\documentstyle[aps,eqsecnum,epsfig]{revtex}
\tighten

\begin{document}
\twocolumn[\columnwidth\textwidth\csname@twocolumnfalse\endcsname

\draft
\title{Shape Coexistence and the Effective
Nucleon-Nucleon Interaction}

\author  {P.-G. Reinhard,$^{1,2}$
D.J. Dean,$^{3,4}$
W. Nazarewicz,$^{3,4,5}$
J. Dobaczewski,$^{1,3,5}$
J.A. Maruhn,$^{1,6}$ and
M.R. Strayer$^{3,4}$
}

\address {$^1$Joint Institute for Heavy Ion Research,
              Oak Ridge National Laboratory,
              P.O. Box 2008, Oak Ridge,   Tennessee 37831}
\address {$^2$Institut f\"ur Theoretische Physik, Universit\"at Erlangen
              Staudtstr.\ 7, D-91058 Erlangen, Germany}
\address {$^3$Department of Physics and Astronomy, University
              of Tennessee Knoxville,  Tennessee 37996}
\address {$^4$Physics Division, Oak Ridge National Laboratory,
              P.O. Box 2008, Oak Ridge,   Tennessee 37831}
\address {$^5$Institute of Theoretical Physics, Warsaw University,
              ul. Ho\.za 69, PL-00681, Warsaw, Poland}
\address {$^6$Institut f\"ur Theoretische Physik, Universit\"at Frankfurt,
              Robert-Mayer-Str. 10, D-60054 Frankfurt, Germany}

\maketitle
\begin{abstract}
The phenomenon of shape coexistence is discussed within the
self-consistent Hartree-Fock method
and the nuclear shell model.
The occurrence of the coexisting configurations with
different intrinsic shapes is
traced back to the properties of the effective
Hamiltonian.
\end{abstract}


\addvspace{5mm}]

\narrowtext

\section{Introduction}\label{intro}

The phenomenon of nuclear  coexistence manifests itself in the presence
of close-lying nuclear states with very different intrinsic properties.
Spectacular examples of coexistence are
superdeformed  states, low-lying deformed states in spherical nuclei,
 high-$K$ isomers, and  pairing
isomers.
One of the most exciting aspects of
the coexistence phenomenon  is the fact
that the  coexisting excited states often retain their identity at
 rather high excitation energies.
Indeed, the lowest observed gamma transitions in
superdeformed  bands
correspond to states lying several MeV above the yrast line, i.e., in the
region of very high level density. This means that
superdeformed configurations
do not mix with many near-lying states; they are very diabatic.

In most cases,
 energies of coexisting states strongly depend
on particle numbers. For instance, the
neutron-deficient Hg isotopes
have well-deformed prolate ground states containing the high-$j$ proton orbitals
``intruding" across the $Z$=82 gap while the ground states of heavier
Hg isotopes are only weakly deformed and they can be associated with oblate shapes.
Such crossings between coexisting structures are particularly interesting in
light nuclei; they can give rise to the presence of deformed ground
states in magic nuclei such as $^{32}$Mg. For an extensive review
of shape coexistence in light and heavy nuclei, we would like to refer the
reader to Refs.~\cite{[Hey83],[Woo92]}.

In the mean-field approach, the
mean fields, in  which nucleons move as independent
(quasi)particles, can be
obtained from a knowledge of the effective forces acting
between nucleons using the Hartree-Fock (HF)
theory.  For a particular choice of nucleon-nucleon force and proton and
neutron numbers, the mean  field may be spherical or deformed.
The minimization of the
HF energy under the constraint of holding any given nuclear multipole moment
fixed can be  carried out over a range 
of collective parameters.  When more than one
local minimum occurs for the total energy as a function of deformation,
shape coexistence may result.

The phenomenon of shape coexistence can be traced back
to the nuclear shell structure. First, the sizes of
spherical magic and
semi-magic gaps in the energy spectrum determine the relative
positions of many-particle many-hole intruder configurations with respect
to the ground state. Secondly, the strength of the particle-vibration coupling
responsible for the breaking of spherical symmetry (i.e., the
development of deformation) is strongly dependent on the
relative distance  between individual shells \cite{[Rei84],[Naz94a]}.
This can be qualitatively understood by means of
Strutinski energy theorem \cite{[Str68],[Bra72],[Str74]}, which states that the
total HF energy can be written as
\begin{equation}\label{strut}
E^{\rm HF}_0 = \tilde{E} +  E_{\rm shell}  + O(\delta\rho^2),
\end{equation}
where $\tilde{E}$ is the smoothed HF energy,
\begin{equation}\label{shell}
 E_{\rm shell}={\rm Tr}{h}^{\rm HF}\delta\rho
\end{equation}
is the shell energy (shell correction),
${h}^{\rm HF}$ is the
HF Hamiltonian, $\rho$ is the single-particle density, and
$\delta\rho$ represents the contribution to $\rho$ due to shell effects.
The shell energy reflects  single-particle properties
of the Hamiltonian; it lowers the binding energy if the Fermi level is
situated in the region of low level density  \cite{[Bra72]}. Hence,
$E_{\rm shell}$ is very sensitive to the
details of the  single-particle spectrum.
 The average HF energy
 $\tilde{E}$ behaves, roughly,  like that of the nuclear liquid drop
\cite{[Bra85]}. In contrast to the shell correction term,
it reflects the average properties of the interaction. In the geometric picture,
deformation comes from the subtle interplay between $\tilde{E}$ and
$E_{\rm shell}$.

It is important to realize that,
 in the mean field approach, the excitation energy
of the coexisting (intruder) state
involves a difference between the binding energies
of coexisting minima, hence
it can easily be obscured by a different
quality of the theoretical description for
these states.
For instance, in  recent work \cite{[Hee98]}, the self-consistent
mean-field theory was used
to explain at the same time the two-particle separation energies in
the first and second wells, and the excitation energies of
superdeformed states.
While for the particle separation energies in the first and
second wells
a good agreement with experimental data could be found, this did not hold
for  {\em relative} energy differences between the
wells. This example demonstrates  that
while the  intrinsic configurations of coexisting states do not
depend in a very sensitive way on details of calculations,
the relative energies of coexisting states
 are strongly affected by  model uncertainties such
as  treatment of pairing, surface tension, or level density, i.e.,
those particular properties of  force
parameterizations
that determine the deformability of a  nucleus.

Another source of
uncertainty in HF calculations
concerns the corrections which should be added to
the calculated energies to account for dynamical
correlations associated  with zero-point fluctuations.
Since the dynamical corrections  can be different in coexisting minima,
they  can influence the predicted excitation energies
\cite{[Cam75],[Rei78],[Rei87]}.

Thus far, we have discussed shape coexistence from the
viewpoint of the  mean-field theory. An alternative, and in
many respects complementary, approach is provided by the nuclear 
shell-model calculations which aim to determine the fully correlated states
in an expansion basis of the few active shells near the Fermi energy.
Here also,
the interplay of the shell structure in nuclei and the
deformation-driving  proton-neutron residual interaction
is the key to understanding the shape coexistence in terms of the
spherical shell model.
The main mechanism for shape coexistence here
is the multiple particle-hole (or pair)
excitation across the closed shell \cite{[Woo92]}.
 By
promoting nucleons out of the closed core into the next higher
shell, nucleons of both kinds  interact to
produce deformed structures.
This approach to shape coexistence was first invoked \cite{[Mor56]}
to explain the 0$^+$ state at 6.05 MeV in $^{16}$O.

In the shell-model framework, the band-head of the
deformed intruder configuration can be written
as \cite{[Hey87]}:
\begin{equation}\label{heyde}
E^*= \Delta E_{\rm ph}
+ \Delta E_{\rm pair} + \Delta E_M + \Delta E_Q,
\end{equation}
where the contributions to the intruding coexisting
 state energy are: $\Delta E_{\rm ph}$ --
the unperturbed particle-hole excitation energy; $\Delta E_{\rm pair}$
-- the change in  pairing correlation energy resulting from the particle-hole
excitation; $\Delta E_M$ -- the change in (proton-neutron)
monopole interaction energy;
and $\Delta E_Q$ -- the change in
(proton-neutron) quadrupole interaction energy.
In Ref. \cite{[Hey87]}, the $\Delta E_{\rm ph}$
and  $\Delta E_{\rm pair}$ terms were estimated using experimental
separation energies, and
 the $\Delta E_M$ and $\Delta E_Q$
terms were calculated using a delta
interaction
for $\Delta E_M$ and the schematic quadrupole-quadrupole
interaction for $\Delta E_Q$.
When applying Eq.~(\ref{heyde}) to  pair excitations in
closed-shell nuclei, it was found that the pairing energy,
monopole energy, and quadrupole interaction energy
tend to  reduce
$\Delta E_{\rm ph}$. For a fixed value of $N$,
the monopole energy monotonically decreases with $Z$,
and  the quadrupole term reaches its maximum
at midshell.

The main objective of this paper is to trace  the phenomenon
of shape coexistence to  properties of the effective nuclear
Hamiltonian.
In our analysis, we apply the mean-field self-consistent
Skyrme-HF method
and the nuclear shell model. The
sensitivity of the
interplay between the
coexisting configurations is discussed in terms of several key
quantities such as the single-particle splitting, pairing correlations,
and surface tension (in the mean-field approach), as well as the
monopole energy, quadrupole correlation energy, and
effective single-particle spectra (in the shell model).

The paper is organized as follows.  The four regions of shape
coexistence studied in this work,
namely the deformed $N$=20 and $N$=28 regions, the
region around $^{80}$Zr, and the region around $^{98}$Zr, are
briefly reviewed in Sec.~\ref{examples}.  The results of the mean-field
and shell-model analysis are contained in Secs.~\ref{HF} and
\ref{SM}, respectively.  Finally, the main conclusions of this work
are summarized  in
Sec.~\ref{conclusions}.

\section{Examples of shape coexistence}\label{examples}

\subsection{The deformed $N$$\approx$20 region}\label{32Mg}

The neutron-rich nuclei with $N$$\approx$20
are spectacular
examples of coexistence between spherical and deformed
configurations in the $sd$ shell (8$\le$ $Z,N$ $\le$ 20).
A classic example  is the magic nucleus $^{32}_{20}$Mg, which
has a very low-lying 2$^+$ state at 886 keV \cite{[Det79]} and
an anomalously high value of
the two-neutron separation energy $S_{2n}$.  Ground-state
deformations in this mass region can also explain an anomalous
isotope shift in $^{31}$Na \cite{[Tou82]} and a major decrease in $S_{2n}$ in
$^{31,33,35}$Na and $^{30}$Ne \cite{[Tou82],[Thi75]}.
The large deformation of $^{32}$Mg has been inferred from
the  intermediate-energy Coulomb excitation studies \cite{[Mot95]}.

In many calculations  based on the  mean-field theory, deformed
ground states  have been predicted in nuclei from the $^{32}$Mg
region (sometimes dubbed as  an ``island of
inversion").  In the early Skyrme-HF  calculations of Ref.
\cite{[Cam75]} with the SIII and SIV interactions,
large prolate deformations in $^{31,32}$Na were obtained  and
explained in terms of neutron excitations from the $sd$
shell  to the $f_{7/2}$
shell.
Similarly, low-lying 2$^+$ states have been predicted
in $N$=20 nuclei based on the energy density formalism \cite{[Bar78a]}.
A sudden onset of large ground-state deformations
($\beta_2$$\sim$0.3--0.4) around $N$=20 was also predicted in the
calculations based on the macroscopic-microscopic method
\cite{[Mol81a],[Ben84]}.

In the shell-model language,
the structural changes around $^{32}$Mg can be attributed to the
cross-shell particle-hole excitations to the $f_{7/2}$ shell.
Early shell-model calculations in a rather restricted
configuration space (no more than two neutrons in the $f_{7/2}$
shell) \cite{[Wat81]} were able to reproduce the increased
quadrupole collectivity at $N$=20.  A similar conclusion
was drawn in other shell-model calculations in the ($sdpf$)
model space \cite{[Pov87],[War90a],[Fuk92],[Pov94],[Cau98]}, 
allowing only two-neutron particle-hole excitations from the $sd$ to the
$pf$ shell,
and in the
schematic analysis of Ref. \cite{[Hey91a]} based on Eq.~(\ref{heyde}).

Recently, the onset of deformations in this region has been a
subject of much  theoretical  work, strongly motivated by the
prospects of detailed experimental spectroscopic studies at ISOLDE
\cite{[Hab97]}. In Ref. \cite{[Ren96]} (see also Ref.
\cite{[Lal98]}), based on the RMF theory with the NL-SH force
and the constant gap BCS treatment of pairing, $^{32}$Mg was
calculated to be spherical. They concluded that this result,
together with the previous RMF study with the NL1 force
\cite{[Pat91a]}, did not depend on the
choice of the  RMF parameterization.
Spherical ground-state deformation for $^{32}$Mg has also been
obtained in the Skyrme-HFB calculations with the SIII, SLy4, and
SkP forces and with a density-dependent zero-range pairing
interaction \cite{[Ter97a]},  in the HFB study based on a
Brueckner $G$-matrix derived from a meson-exchange potential
with the density-dependent meson masses \cite{[Gru96]}, and also
in the Gogny-HFB calculations of Ref.~\cite{[Ber92]}. The authors
of Ref.~\cite{[Ber92]} noted, however, that strong deformation effects
around $^{32}$Mg could appear due to dynamical correlations. Their
 collective wave functions of $^{30}$Ne and $^{32}$Mg,
calculated with the collective Hamiltonian, have
 pronounced maxima
at  large deformations.

\subsection{The deformed $N$$\approx$28 region}\label{44S}

Another, recently discovered, island of inversion are the
neutron-rich nuclei from the $pf$ shell centered around
$^{44}$S$_{28}$. Experimentally, $\beta$-decay properties of
$^{44}$S and $^{45-47}$Cl have been studied in
Refs.~\cite{[Lew89],[Sor93]}. Based on the QRPA analysis of
measured half-lives, it was concluded that $^{44}$S was
deformed.  This has been confirmed recently in a series of
intermediate-energy Coulomb excitation studies
\cite{[Sch96],[Gla97]} which revealed rather large
$B(E2;0^+_{\rm g.s.}\rightarrow 2^+_1)$ values in this
neutron-rich  region, suggesting a significant breaking of the
$N$=28 core.  (For  recent mass measurements around $^{44}$S,
see Ref.~\cite{[Sav98]}.)

The  HF+BCS calculations with the  Skyrme interactions SIII and
SkM$^*$ and the RMF calculations with the  parameter set NL-SH
\cite{[Wer94a],[Wer96]} predicted the
appreciable breaking of the $N$=28
core and deformation effects around $^{44}$S.
A later study
\cite{[Hir96]}, based on the RMF approach without pairing and using
the TM1 parameter set, predicted the neutron-rich sulfur isotopes
to be deformed. In particular, $^{44}$S was found to
be prolate, in agreement with the RMF results of
Refs.~\cite{[Wer94a],[Wer96]}. Only very recently, the interplay
between deformed mean-field and pairing correlations in this
mass region has been properly considered in the framework of
relativistic Hartree-Bogolyubov (RHB) theory \cite{[Lal98a]}
using the NL3 effective
interaction  for
the mean-field Lagrangian
 and the Gogny interaction D1S in the pairing channel.
Again,  deformed shapes around $^{44}$S have been calculated.

An erosion of the $N$=28 gap in the sulfur isotopes has also
been found in shell-model calculations \cite{[Ret97]} performed
in a large configuration space (the full $sd$ shell for protons
and the full $pf$ shell for neutrons). The authors concluded,
however, that the shell-breaking effects around $^{44}$S were
much weaker compared to the $N$$\approx$20 neutron-rich
region.

\subsection{The $N$$\approx$$Z$$\approx$40 region}\label{80Zr}

The proton-rich $N$=$Z$ nucleus $^{80}$Zr
lies in the center of the well-deformed $A$$\approx$80 region
\cite{[Naz85]}.
The  sizable energy gap at
particle number 40 separates the $pf$
spherical shell from the $g_{9/2}$ orbital.
However, this spherical subshell closure
is not sufficiently large
to stabilize the
spherical shape. Experimentally \cite{[Lis87]}  $^{80}$Zr
seems to be a well-deformed rotor. According to the
 mean-field theory, this is due to the presence of
the deformed single-particle gap at $N$,$Z$=40; the resulting
deformed shell effect turns out to be stronger than that
at the spherical shape.

Microscopic  calculations based on the symmetry-projected
variational model \cite{[Pet96]}, Skyrme-HF theory
\cite{[Bon85],[Bon91a]}, RMF theory \cite{[Lal95]}, and
the restricted-space HFB calculations \cite{[Kir93]} predict
a deformed ground-state minimum for $^{80}$Zr.
Only in a very few calculations, such as
the  RMF calculations with the NL1 parameter set \cite{[Mah92]},
was a spherical ground state  obtained.  (See, however, the
discussion in Ref. \cite{[Lal95]}.)

\subsection{The $N$$\approx$56, $Z$$\approx$40 region}\label{96Zr}

Nuclei from the heavy-Zr region ($Z$$\approx$40, $N$$>$56)
exhibit a wealth of coexistence
phenomena \cite{[Woo92],[Zir88]}.
The strong dependence of observed
spectroscopic properties on
the number of  protons and neutrons
makes the neutron-rich $A$$\approx$100 nuclei  a very
good region  for testing various models.
Theoretically,
strong shape variations in this region may be
attributed to shell effects associated with large spherical
and deformed subshell closures in the
single-particle spectrum \cite{[Ska97]}.

According to calculations based on the mean-field approach, the
occupation of the  $h_{11/2}$ neutron  and $g_{9/2}$ proton
orbitals   is essential for understanding the  deformed
configurations near $^{100}$Zr \cite{[Naz88a]}.  The best
examples of shape coexistence in this region are
the  Sr, Zr, and Mo
isotopes with $N$$\approx$58.  In the language of the deformed
shell model, the onset of deformation around $N$=58 can be
associated with the competition between the spherical gaps at
$Z$=38, 40, and  $N$=56, and the deformed subshell closures at
 particle numbers $Z$=38, 40, and
$N$=60, 62, and 64.  Theoretically, the delicate energy  balance
between spherical and deformed configurations depends crucially
on the size of these gaps.  As
discussed in Refs.~\cite{[Woo92],[Dob88],[Wer94]}, the
deformation onset at $N$$\approx$58 results from  the subtle
interplay between the deformation-driving neutron-proton 
quadrupole interaction
and the
symmetry-restoring monopole force  responsible for  shell
effects.

Equilibrium deformations and moments, potential energy surfaces,
the microscopic structure of coexisting
configurations, and shape transitions in the heavy-Zr region
have been calculated by many authors. (For an extensive
list of references, see Ref.~\cite{[Ska97]}.)
In most cases, the  calculations show large deformations in
the Sr, Zr, and Mo isotopes with $N$$\geq$60. The
details of the  shape
transition near  
 $N$=58 are, however,  predicted differently by
various models, the onset and rapidity of this transition
being very
sensitive to the actual parameterization used \cite{[Bon85],[Woo92]}.

\section{Skyrme-Hartree-Fock Calculations}\label{HF}

\subsection{The Skyrme Energy Functional}\label{Skyrme_model}

Our implementation of  Skyrme forces is based on the standard
ansatz as it has now been  used for more than two decades
\cite{[Que78]}.
The total binding energy of a nucleus is obtained self-consistently
from the  energy functional:
\begin{eqnarray}
   {\cal E}  =  {\cal E}_{\rm kin}
             &+&{\cal E}_{Sk}(\rho,\tau)
               +{\cal E}_{Sk,ls}(\rho,\bbox{J}) \nonumber \\
             &+&{\cal E}_{C}(\rho_p)
               +{\cal E}_{C,ex}(\rho_p)
               +{\cal E}_{\rm pair}
               -{\cal E}_{\text{CM}}, \label{eq:Etot}
\end{eqnarray}
where
\begin{eqnarray}
   {\cal E}_{\rm kin} &= &\int d^3r  \frac{\hbar^2}{2m} \tau ,
\label{Ekin} \\
   {\cal E}_{Sk} &= &\int d^3r \left\{
                 \frac{b_0}{2}  \rho^2
          + \frac{b_3}{3}  \rho^{\alpha +2}
     +  b_1 \rho \tau
     -\frac{b_2}{2} \rho \Delta \rho  \right.
            \nonumber \\
     &-& \left. \!\sum_q\! \left(\frac{b'_0}{2} \rho_q^2
          + \frac{b'_3}{3} \rho^\alpha \rho_q^2
     + b'_1 \rho_q \tau_q
      -\frac{b'_2}{2} \rho_q \Delta \rho_q\right)\!\!\right\}\!,\!\!
\label{eq:Esky}  \\
 {\cal E}_C & = & \frac{1}{2}  e^2 \int d^3r \, d^3r' \rho_p(\bbox{r})
              \frac{1}{|\bbox{r}-\bbox{r}'|} \rho_p(\bbox{r}'),
             \label{eq:Ecoul}
\end{eqnarray}
$ {\cal E}_{Sk,ls}$ is the spin-orbit functional,
${\cal E}_{\rm pair}$ is the pairing energy, and
${\cal E}_{\text{CM}}$ is the center-of-mass correction.

The functional employs the usual particle densities
$\rho_q = \sum_{\alpha\in q}n_\alpha|\psi_\alpha|^2$,
the
kinetic densities $\tau_q = \sum_{\alpha\in
q}n_\alpha|\bbox{\nabla}\psi_\alpha|^2$,
and  the spin-orbit densities
$\bbox{J}_q=\sum_{\alpha\in q}
n_\alpha\psi_\alpha^+\bbox{\sigma}\times\bbox{\nabla}\psi_\alpha$,
 where $\psi_\alpha$ are
the single-particle (canonical) wave functions and $q$ stands for either
protons or neutrons. The total
isoscalar density is $\rho$=$\rho_p$+$\rho_n$
and similarly for $\tau$ and $\bbox{J}$. The 
$n_\alpha$=$v_\alpha^2$ is the
BCS occupation weight (see below).

The  terms  discussed above
are always  defined in the same way
for all Skyrme parameterizations. This is not the case for 
 the remaining
terms in  Eq.~(\ref{eq:Etot}). The Coulomb-exchange
functional is usually treated in the Slater approximation
\begin{equation}
 {\cal E}_{C,ex}
   =
         -\frac{3}{4} e^2\left(\frac{3}{\pi}\right)^\frac{1}{3} \int d^3r
                          [ \rho_p(\bbox{r})]^\frac{4}{3},
  \label{Ecoulex}
\end{equation}
but it is omitted in definitions of some published Skyrme forces.
All the parameterizations considered in this work
 require this term.

The spin-orbit functional can be written as
\begin{eqnarray}
   {\cal E}_{Sk,ls}
   &=&
   \int d^3r
     \left\{- b_4 \rho\bbox{\nabla}\cdot\bbox{{J}}
     -b'_4 \sum_q \rho_q(\bbox{\nabla}\cdot\bbox{J}_q)\right.
\nonumber\\
   &&\quad
     +\left.\frac{\theta_{\text{ls}}}{12}
      \left[\left(\frac{3}{2}b_1+b_2-b'_1+6b'_2\right)\bbox{J}^2 \right.\right.
\nonumber\\
   &&\quad
     \left. \left.-\left(b_1+2b_2-\frac{1}{2}b'_1+3b'_2\right)\sum_q\bbox{J}_q^2\right]
     \right\}.
\label{eq:Eskyls}
\end{eqnarray}
This spin-orbit functional encompasses two different options,
namely, one either
ignores  the $\bbox{J}^2$ contributions ($\theta_{\text{ls}}=0$)
or  takes them into account ($\theta_{\text{ls}}=1$).
 Furthermore,
the spin-orbit functional (\ref{eq:Eskyls}) is
given in the extended form of \cite{[Rei95]} which allows a
separate adjustment of isoscalar and isovector spin-orbit
force.  The standard Skyrme forces use the particular
combination $b'_4$=$b_4$ which was motivated by the derivation
from a two-body zero-range spin-orbit interaction \cite{[Vau72]},
 but these particular settings are not obligatory when
taking the viewpoint of an energy-density functional. Thus,
various options exist in the published literature, and we shall
use all combinations of them in the examples discussed in this work.

Similarly, there are basically two different options for handling the
center-of-mass correction, i.e.,
\begin{eqnarray}
      {\cal  E}_{\text{CM}}
      &=&
      \theta_{\text{CM}}\frac{\hbar^2}{2m} \langle\hat{P}^2_{\text{CM}} \rangle
\nonumber \\
      &+&(1-\theta_{\text{CM}})\frac{\hbar^2}{2m(A-1)}
      \sum_\alpha n_\alpha
\langle \psi_\alpha|\hat{p}^2|\psi_\alpha\rangle .
\label{eq:Ecm}
\end{eqnarray}
For $\theta_{\text{CM}}$=0, the center-of-mass correction
is implemented {\em before variation} by the simple trick
to let the nucleon mass $m\rightarrow m_{\rm red}\equiv m-m/A$. The
option $\theta_{\text{CM}}$=1 uses a more correct expression, but it is
difficult to implement in the fully self-consistent manner
  due to the two-body nature of the $\hat{P}^2_{\text{CM}}$ operator.
 Hence standard
parameterizations with $\theta_{\text{CM}}$=1 apply this correction
{\em after variation} for
the given mean-field solutions obtained with the center-of-mass correction
ignored.
(For examples of a fully self-consistent treatment, see Ref.~\cite{[Cha95a]}.)

Since there are more than  80 different Skyrme parameterizations
on the market,
the question arises, which forces
should actually be  used when making predictions and comparing with the data?
 An extensive  list of forces, together
with their properties, can be found in
Ref.~\cite{[Str98]}. To select a manageable number, we
have computed the overall quality factor which reflects
the predictive power of the force for  the
basic ground-state properties (masses, radii, surface
thicknesses), and confined further analysis to the best
performing parameterizations.
From these, we have chosen a still smaller subset
with sufficiently different properties to explore the
possible variations among parameterizations.  This subset
contains: SkM$^*$ \cite{[Bar82]}, SkT6
\cite{[Ton84]}, Z$_\sigma$  \cite{[Fri86]},
SkP \cite{[Dob84]}, SLy4
\cite{[Cha95a]}, and SkI1, SkI3, and SkI4 from
Ref.~\cite{[Rei95]}. We have also added two
additional  forces from a recent
exploration \cite{[Ben98a]}.
These two forces are labeled SkO and SkO'.
A  list of the parameters for these forces is given in 
Appendix~\ref{sec:appforces}.

All the selected forces  perform  well concerning
the total energy and
radii.  They all have comparable incompressibity
$K$=210-250\,MeV and comparable surface energy
which  results from a
careful fit to ground-state properties. Variations
occur for properties which are not
fixed precisely by ground-state characteristics. The effective
nucleon mass is 1 for SkT6 and SkP,
0.9 for SkO and SkO',
around 0.8 for SkM$^*$ and Z$_\sigma$, and even lower,
around 0.65, for SLy4, SkI1, SkI3, and SkI4.
Isovector properties  also exhibit large variations.
The asymmetry energy ranges from very low, 26\,MeV for  Z$_\sigma$,
to rather high, 38\,MeV for SkI1, with the values for other forces
being around 30-32\,MeV. The appropriate options for
the center-of-mass correction (\ref{eq:Ecm}) and spin-orbit force
(\ref{eq:Eskyls}) are found in Table~\ref{tab:forces} in
Appendix~\ref{sec:appforces}. The choice embraces
 $\theta_{\text{CM}}$=0 as well as $\theta_{\text{CM}}$=1,
and the various options for the spin-orbit force. 
The only forces in the sample which have not yet been published
elsewhere are SkO and SkO'. They stem from an ongoing exploration
of Skyrme forces trying to accommodate more observables. In addition to
SkI4 which fits ground-state energies, radii, surface thicknesses and
the isotope shifts of r.m.s. radii, SkO and SkO' also manage 
to reproduce the jump in the isotopic trend of the two-neutron
separation energies in the lead isotopes, a feature where most Skyrme forces fail.
Moreover, these two forces represent a most recent update of the fits
along the  line of \cite{[Fri86],[Rei95]} now using an up-to-date treatment of
pairing, see Sec.~\ref{sec:pairing}. Last but not least, we 
have here a pair of forces which are fitted in precisely the same
manner and differ only in the spin-orbit factor $\theta_{\rm ls}$.
This allows for testing specifically the impact of this variation.

\subsection{Treatment of pairing}
\label{sec:pairing}

In the original publications, various
forces were  used with different pairing
recipes. Most of these recipes are very schematic
(e.g., constant gap or
seniority force)
and  fail
when proceeding into the regime of exotic nuclei.
(See discussion in Refs. \cite{[Dob84],[Dob96]}.)  On
the other hand, details of the actual pairing recipe
do not affect the overall quality of the forces
because these are usually  fitted to properties
of well-bound nuclei.
In this work,
we compute the pairing matrix elements from a local
interaction.  Among several  choices,
we take the simple $\delta$ force, which
leads
to the pairing energy functional
\begin{equation}
\label{eq:PairFunc}
  {\cal E}^{\rm pair}
  =
  \sum_{q\in\{p,n\}}\frac{V_{q}}{4} \int \! {\rm d}^3r \; \chi_q^2 (\bbox{r})
    F(\bbox{r}),
\end{equation}
where the local pair density $\chi ({\bf r})$ reads \cite{[Dob84],[Dob96]}
\begin{equation}
\chi_q (\bbox{r})
= \sum_{\alpha \in q} u_\alpha \, v_\alpha \,
    |\psi_\alpha (\bbox{r})|^2,
\end{equation}
and the function $F$=1 or $F$=$1-\rho(\bbox{r})/\rho_c$
gives the volume or surface type of pairing correlations, respectively,
while $\rho_c$=0.16\,fm$^{-3}$
 is the saturation density.

A further key quantity for pairing is the selection of
the pairing  phase space. Following Refs.~\cite{[Kri90],[Bon85]},
to cut the space at the high-energy side,
we  use a  Fermi-type form factor
\begin{equation}
f_\alpha = \frac{1}{1 + \exp \left[(\epsilon_\alpha - \lambda - \Delta E)/\mu
\right] },
\label{cutoff}
\end{equation}
where the width of the smooth cut-off is linked to the energy offset
by
\begin{equation}
 \mu
 =
 \frac{\Delta E}{10},
\end{equation}
as  was done in the earlier proposals \cite{[Kri90],[Bon85]}.
We adopt the
point of view that pairing is a valence-particle effect
 for which the energy range
$\Delta E$ should be proportional to the average
level spacing for the given nucleus and nucleon type.
We accomplish this
 by fixing the number of pairing-active states:
\begin{equation}\label{nact}
  N_{\rm act}
  =
  2\sum_\alpha f_\alpha =  {\cal N} + 1.65\,{\cal N}^{2/3},
\end{equation}
where ${\cal N}$=$Z$ or $N$.
 Details of this ``soft" cut-off scheme and the reasoning behind
the actual choice of the factor 1.65 in Eq.~(\ref{nact})
can be found in
Ref.~\cite{[Ben98]}.

Having selected the pairing recipe, one needs  to
fix the strengths $V_{p}$ and
$V_{n}$.
They have been fitted to  empirical pairing
gaps in a selection of nuclei.
Details and the 
resulting  pairing force parameters  are given
in Table~\ref{tab:forces} in
Appendix~\ref{sec:appforces}. 
The surface-pairing strengths for neutrons and protons
have been adjusted in the same way as the strength
parameters of the standard (volume) delta pairing.

\subsection{Description of calculated quantities}\label{observables}

After solving  the
HF equations in the usual manner, we obtain the self-consistent
single-particle orbitals from which the total energy, as well
as several other observables, can be calculated, as described in this section.

A global characteristic of
the Skyrme interaction  is the surface energy coefficient:
\begin{equation}\label{surface}
  a_s
  =
  \lim_{A\rightarrow\infty}
  \left[\frac{E(A)}{A}-\left(\frac{E}{A}\right)
   \mbox{\rule[-1em]{.1mm}{2.5em}}_{~A=\infty}\right]A^{1/3},
\end{equation}
where $E(A)$ is the energy computed for $A$ particles. Because
$a_s$ is computed from the HF results for large particle numbers, it
is independent of shell effects, and hence it characterizes
the surface properties of the bulk energy $\tilde{E}$ of Eq.~(\ref{strut}).
As the limiting process in Eq.~(\ref{surface})
is extremely slow \cite{[Tre81]}, it is best to
evaluate $a_s$  for semi-infinite nuclear matter, and  for
that we use the semiclassical  M. Brack code~\cite{[Bra85]}.

\begin{figure}
\begin{center}
\leavevmode
\epsfxsize=16cm
\epsfbox{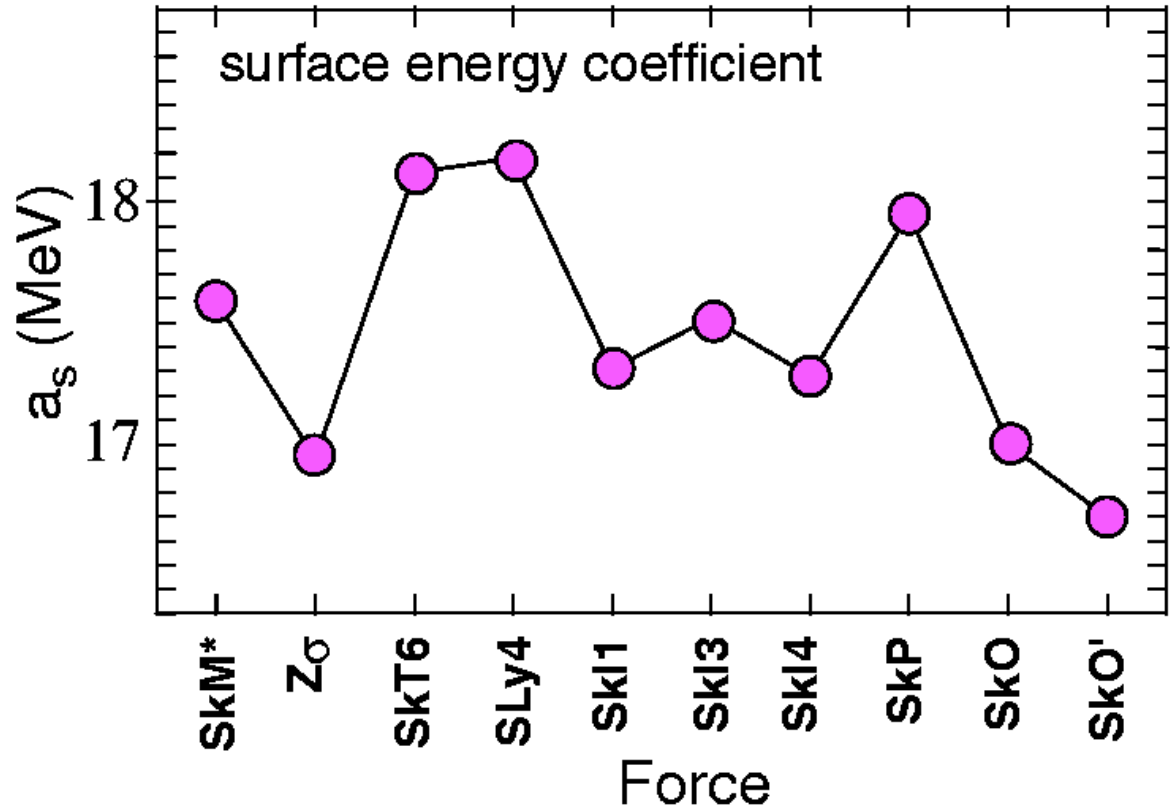}
\end{center}
\caption{The surface energy coefficient as defined in
Eq.~(\protect\ref{surface}) for the  Skyrme
parameterizations used in this work.
}
\label{asurf}
\end{figure}

Figure~\ref{asurf}
displays the  surface energy coefficient for
the Skyrme parameterizations employed  in this work. The larger the $a_s$,
the greater the surface tension. Consequently,   large  values
of $a_s$ imply the stronger resistance of the system against surface distortions
(or, in other words, reduced deformability). As
seen in Fig.~\ref{asurf},
the ``stiffest" interactions
are SkT6, SLy4, and SkP, and
the ``softest" parameterizations are Z$_\sigma$, SkO, and SkO'.

A large part of our survey below deals
with quadrupole deformation potentials.
We produce a systematic series of deformed mean-field states
by adding a quadrupole constraint
$\hat{Q}\propto r^2Y_{20}/[1+f(r)]$ to the HF field,
where the function $f(r)$ suppresses
$\hat{Q}$  at large distances (see Ref.~\cite{[Fin89]}).
The calculated deformed shapes are characterized by
means of the dimensionless
quadrupole deformation:
\begin{equation}
  \beta
  =
  \sqrt{\frac{\pi}{5}}
  \frac{\langle r^2 Y_{20}\rangle}{A\langle r^2\rangle}.
\end{equation}
The total energy of the system $E_{\rm tot}$ as a function of $\beta$
represents a zero-order approximation to
the potential energy curve for $\beta$-vibrations, i.e.,
\begin{equation}
  {\cal V}(\beta)
  \equiv
  E_{\rm tot}(\beta)
  =
  \langle\Phi_{\beta}|\hat{H}|\Phi_{\beta}\rangle.
\end{equation}
However, before one can use
${\cal V}(\beta)$
 in calculations with the
collective Hamiltonian, dynamical corrections
 have to be added. The reason is that the underlying states
$|\Phi_{\beta}\rangle$ have a finite uncertainty in the collective
deformation, i.e., $\Delta^2\beta\neq 0$. As a consequence, the
potential ${\cal V}(\beta)$ contains contributions from
$\beta$-fluctuations in $|\Phi_{\beta}\rangle$,
 and these contributions need to be subtracted
first before adding the energies associated with the true physical
zero-point fluctuations in $\beta$. The theoretical evaluation of
these correction terms can be done in the framework of the Generator
Coordinate Method  at the level of the Gaussian Overlap
Approximation (GOA), as has been discussed in several publications.
(See Ref. \cite{[Rei87]} for a review.) The collective parameters in the
present (axially symmetric) case are the  quadrupole deformation
$\beta$ and the two rotational angles $\vartheta$ and $\varphi$.
The volume element in these coordinates is not Cartesian and thus
one has to employ the GOA in a topologically invariant fashion.
For a detailed discussion of the general case, see
Ref.~\cite{[Goz85]}. Simpler formulae used in this work  are taken from
Ref.~\cite{[Rei78]}, namely,
we  define the corrected deformation energy as
\begin{equation}\label{eq:ZPEcor}
  {V}(\beta)
  =
  {\cal V}(\beta)
  -
  \left(E_{{\rm ZPE},\beta}+E_{\rm ZPE,rot}\right), \label{Vbeta}
\end{equation}
where the rotational, $E_{\rm ZPE,rot}$, and vibrational,
$E_{{\rm ZPE},\beta}$, zero-point energy
corrections read
\begin{eqnarray}
  E_{\rm ZPE,rot}
  &=&
  \frac{\langle\hat{\bbox{I}}^2\rangle}{2\Theta_{\rm rot}}\label{ZPErot},
\\
  E_{{\rm ZPE},\beta}
  &=&
  \frac{2\langle\stackrel{\leftarrow}{\partial_{\beta}}
                \!\hat{H}\!\stackrel{\rightarrow}{\partial_{\beta}}\rangle
        \!-\!\langle\stackrel{\leftarrow}{\partial}_{\beta}^2\!\hat{H}\rangle
        \!-\!\langle\hat{H}\!\stackrel{\rightarrow}{\partial}_{\beta}^2\rangle}
       {8\langle\stackrel{\leftarrow}{\partial_{\beta}}
         \stackrel{\rightarrow}{\partial_{\beta}}\rangle} \!
   \left[3\!-\!g\left(\!\frac{\langle\hat{I}^2_x\rangle}{2}\!\right)\!\right]
\nonumber\\ &&
  - E_{\rm ZPE,rot}\left[1-g\left(\!\frac{\langle\hat{I}^2_x\rangle}{2}\!\right)\right]
.\label{ZPEvib}
\end{eqnarray}
The rotational moment of inertia is determined from
\begin{equation}
  \frac{1}{\Theta_{\rm rot}} =
  \langle[\hat{I}_x,[\hat{H},\hat{I}_x]]\rangle, \label{MOI}
\end{equation}
while the switching factor $g\left(\langle\hat{I}^2_x\rangle/2\right)$,
which originates from the topologically
invariant extension of GOA \cite{[Rei78]},
is defined as
\begin{equation}
  g(\eta) =
  \frac{\int_0^1 dx\, \eta(x^2-1)e^{\eta(x^2-1)}}
       {\int_0^1 dx\, e^{\eta(x^2-1)}}.
\end{equation}
In Eqs.~(\ref{ZPErot})--(\ref{MOI}), the average values $\langle\rangle$ are
taken with respect to the $\beta$-dependent HF states 
$|\Phi_{\beta}\rangle$.
The definition of the collective mass parameters recurs, in principle, to
the full Hamiltonian $\hat{H}$. However, for the present exploratory purposes,
 we employ
the Inglis cranking approximation which is obtained from the above
expressions by letting $\hat{H}\rightarrow\hat{h}_0$, with $\hat{h}_0$ being
the mean-field Hamiltonian.
In the following, the results of calculations of the potential
energy surfaces (PES)  always pertain to the total energies corrected
for the zero-point motion, as in Eq.~(\ref{Vbeta}).

\subsection{Discussion of Potential Energy Surfaces}\label{HF_res}

For the set of
Skyrme parameterizations
described in Sec.~\ref{Skyrme_model},
the PESs
have been calculated
for $^{26,28,30,32}$Ne, $^{30,32,34}$Mg,
$^{38,40,42,44}$S, $^{80,82,84}$Zr, and $^{92,94,96,98,100}$Zr
as functions of quadrupole deformation $\beta$.
These results are discussed below.

\subsubsection{Deformation in the $N$$\approx$20 region}

\begin{figure}
\begin{center}
\leavevmode
\epsfxsize=16cm
\epsfbox{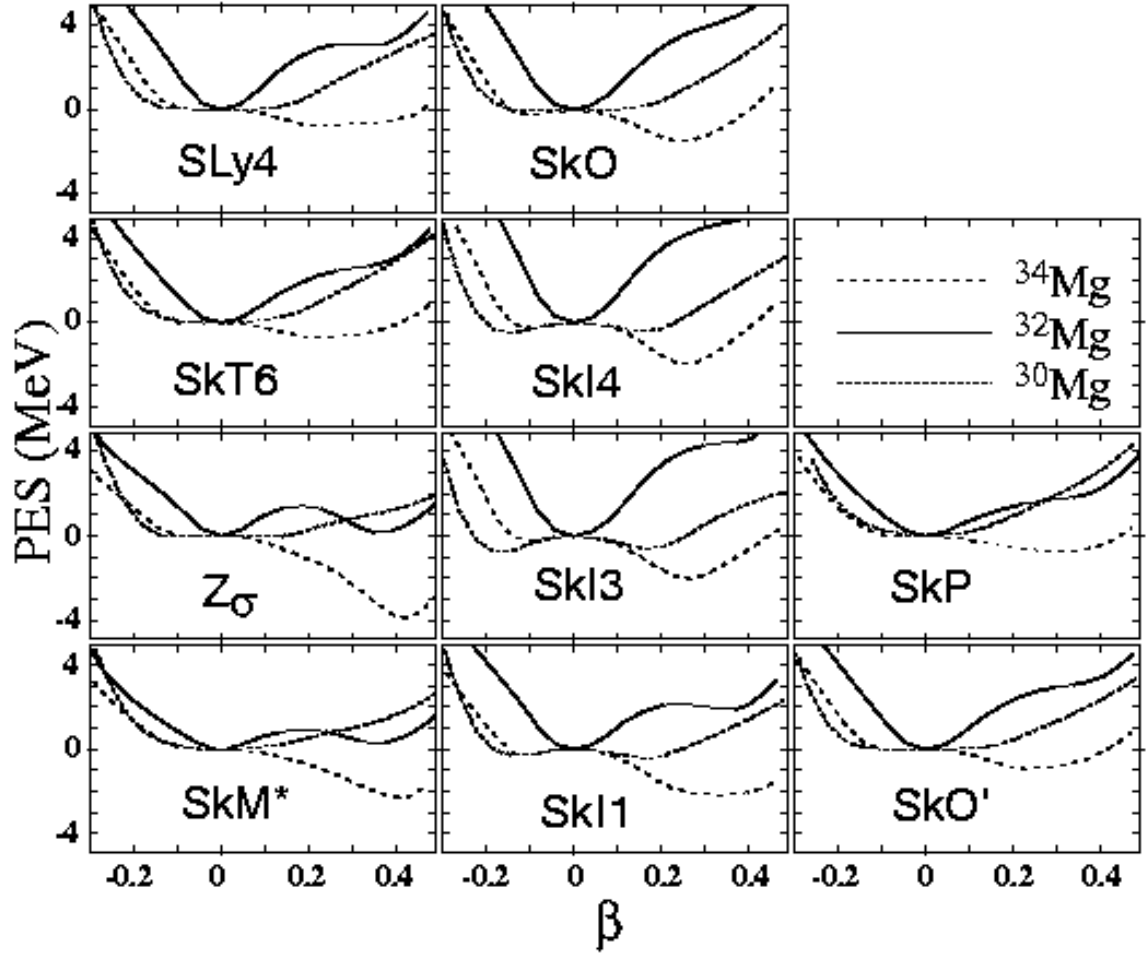}
\end{center}
\caption{Potential energy surfaces for $^{30,32,34}$Mg
as functions of quadrupole deformation $\beta$ for the set
of Skyrme parameterizations.
}
\label{PES-Mg}
\end{figure}

The results of calculations for  $^{30,32,34}$Mg are
shown in Fig.~\ref{PES-Mg}.
For most Skyrme parameterizations
 used, the pattern is fairly
similar. Namely, the nucleus $^{30}$Mg is predicted to be merely
deformation-soft, while the occupation of the  $f_{7/2}$ neutron
shell in $^{34}$Mg
gives rise to a 
very deformed intrinsic shape with $\beta$ ranging from
0.3 to 0.4.  The nucleus  $^{32}$Mg appears to be a transitional system with
coexisting spherical and prolate minima. For  Skyrme parameterizations
SkM$^*$ and $Z_\sigma$, the prolate minimum is calculated 
 to be practically degenerate
with the spherical one. For the remaining forces, the prolate structure
(sometimes corresponding to a local minimum, sometimes forming a shoulder
in the PES)
lies from 2\,MeV to 4\,MeV above the spherical ground state, depending
on the choice of the Skyrme parameterization.

\begin{figure}
\begin{center}
\leavevmode
\epsfxsize=16cm
\epsfbox{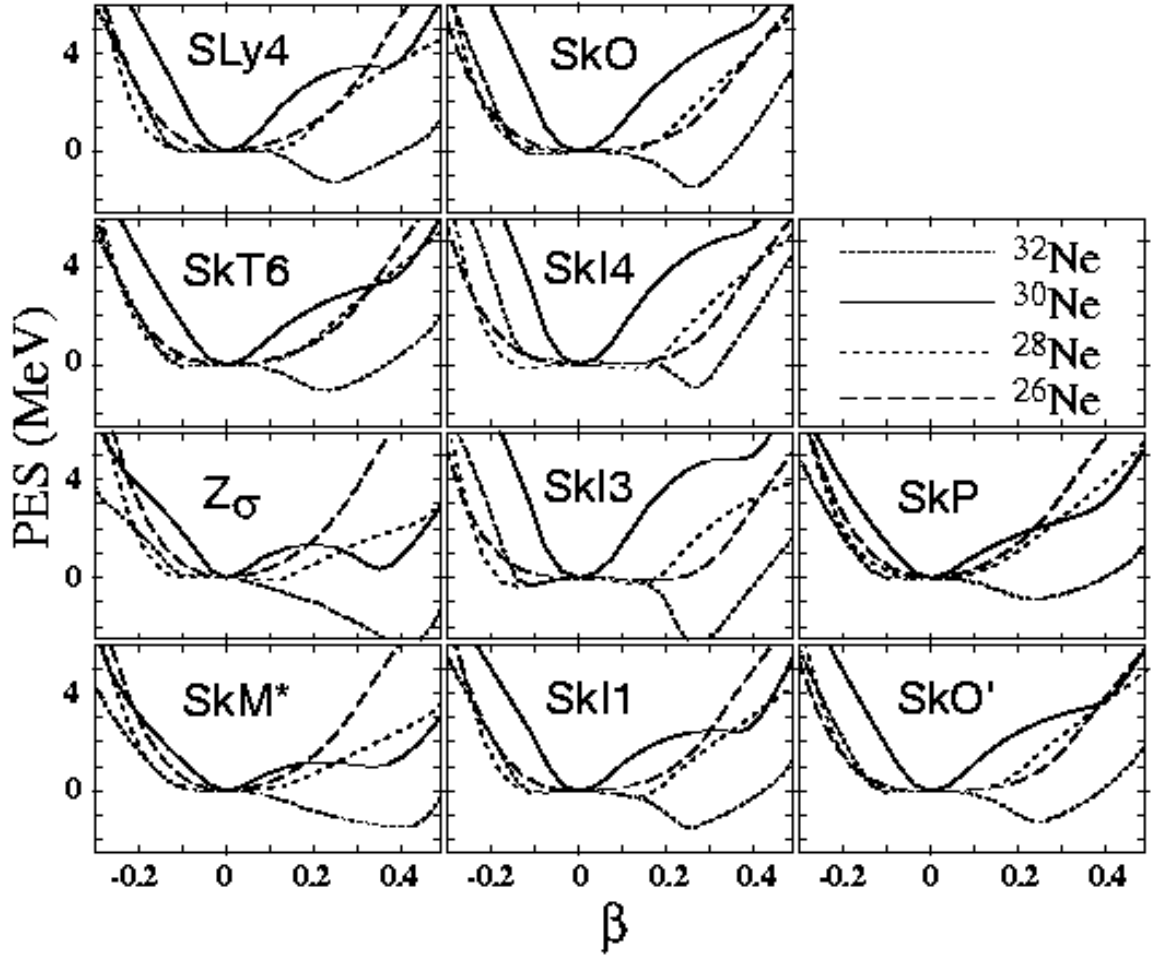}
\end{center}
\caption{Same as in Fig.~\protect\ref{PES-Mg}, except for
 $^{26,28,30,32}$Ne.
}
\label{PES-Ne}
\end{figure}

A similar pattern is observed in
Fig.~\ref{PES-Ne} for the neutron-rich Ne isotopes. Here, the nuclei
$^{26,28}$Ne are predicted to be very soft, strongly anharmonic, while
$^{32}$Ne is well deformed in all cases. The semi-magic $^{30}$Ne is predicted to
be spherical. However, as in the case of  $^{32}$Mg,
a low-lying secondary prolate minimum develops in the
SkM$^*$ and $Z_\sigma$ models. By comparing Figs.~\ref{PES-Mg}
and \ref{PES-Ne} one  notices that for all the forces used,
the deformed configuration in
$^{30}$Ne lies $\sim$1\,MeV higher in energy than that in $^{32}$Mg. That is,
the shape mixing phenomenon
 is expected to be much  stronger in $^{32}$Mg than in $^{30}$Ne.

\begin{figure}
\begin{center}
\leavevmode
\epsfxsize=16cm
\epsfbox{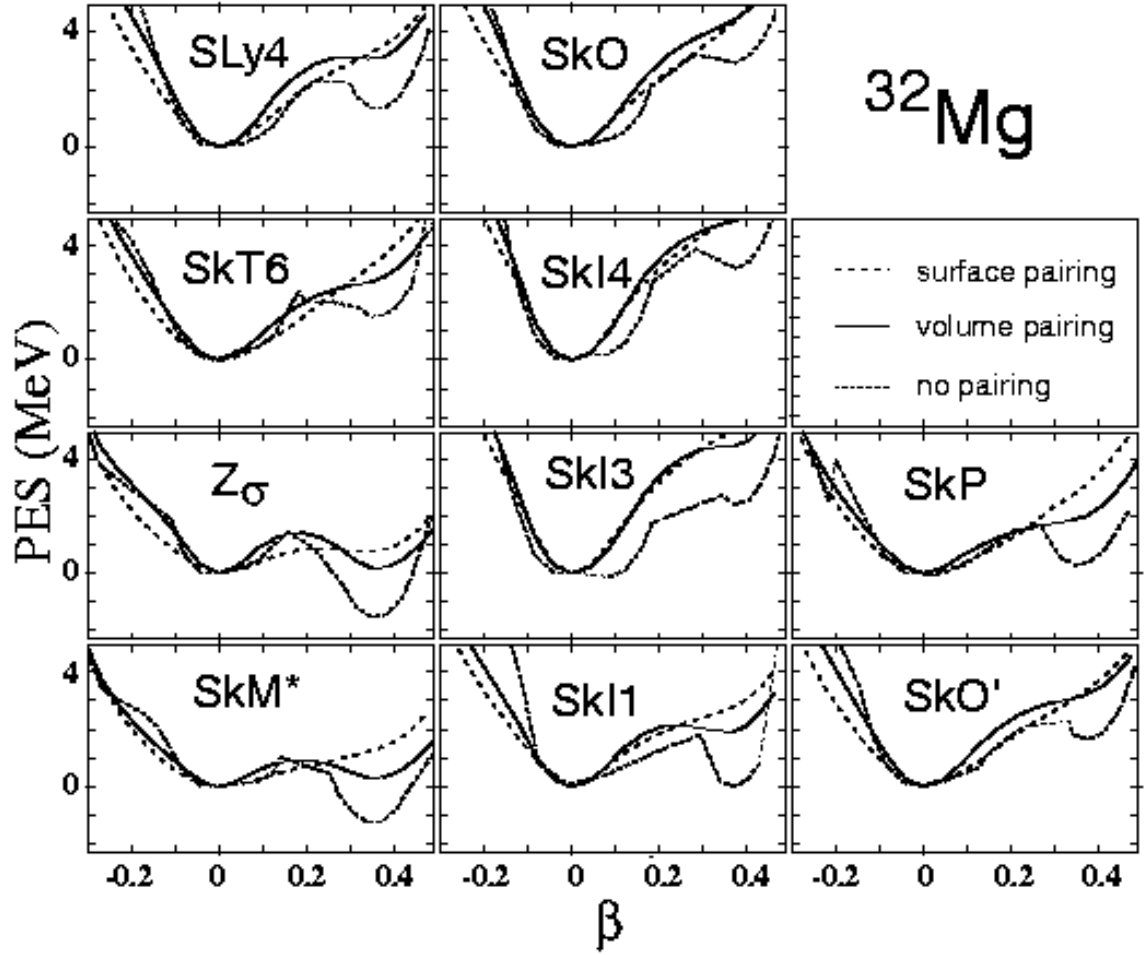}
\end{center}
\caption{Same as in Fig.~\protect\ref{PES-Mg}, except
 for  $^{32}$Mg and for
 three different pairing models: volume delta pairing
(solid line), no pairing (dotted line), and surface pairing
(\protect\ref{eq:PairFunc}) (dashed line).
}
\label{Mg-pairing}
\end{figure}

Of course, in the case of low-lying coexisting states,
the energy difference between spherical and deformed minima
depends strongly on the  details of the calculations.
In particular, variations in the treatment of pairing
correlations are expected to play a role in light nuclei such as
$^{32}$Mg.
 To illustrate
this point, we performed two
additional sets of calculations for $^{32}$Mg using different
pairing recipes.
Figure \ref{Mg-pairing} shows the PESs for $^{30,32,34}$Mg
obtained by taking  (i)  volume pairing as in Fig.~\ref{PES-Mg}, (ii)
the surface pairing interaction
as defined in Eq.~(\ref{eq:PairFunc}),
and (iii)  neglecting  pairing (i.e., pure HF).

As expected, the prolate minimum is well developed in most unpaired
calculations, and its energy is significantly lowered as compared to
the calculations with pairing.
(For the forces SkM$^*$, $Z_\sigma$, and SkI1
  the prolate unpaired minimum becomes the  ground
state.) The opposite holds
for the surface-pairing variant: the corresponding PESs seem softer
 in the direction of  $\beta$.
The sensitivity of the calculated excitation energy of
the intruder state in
$^{32}$Mg on the pairing recipe indicates that
the detailed description would require
 (i) a realistic pairing interaction
that could be applied in mean-field calculations for
light nuclei, and (ii) the proper treatment of
particle-number fluctuations.
Other  uncertainties
in determining the relative energies of coexisting states
are discussed in Sec.~\ref{sec:ZPE} below.

There are many factors that can influence
the  energy  difference between  coexisting states. Probably the most
important one is the  single-particle shell structure. 
Positions of individual shells
are strongly affected by changes in Skyrme parameters, in particular
those defining  the spin-orbit term.

\begin{figure}
\begin{center}
\leavevmode
\epsfxsize=16cm
\epsfbox{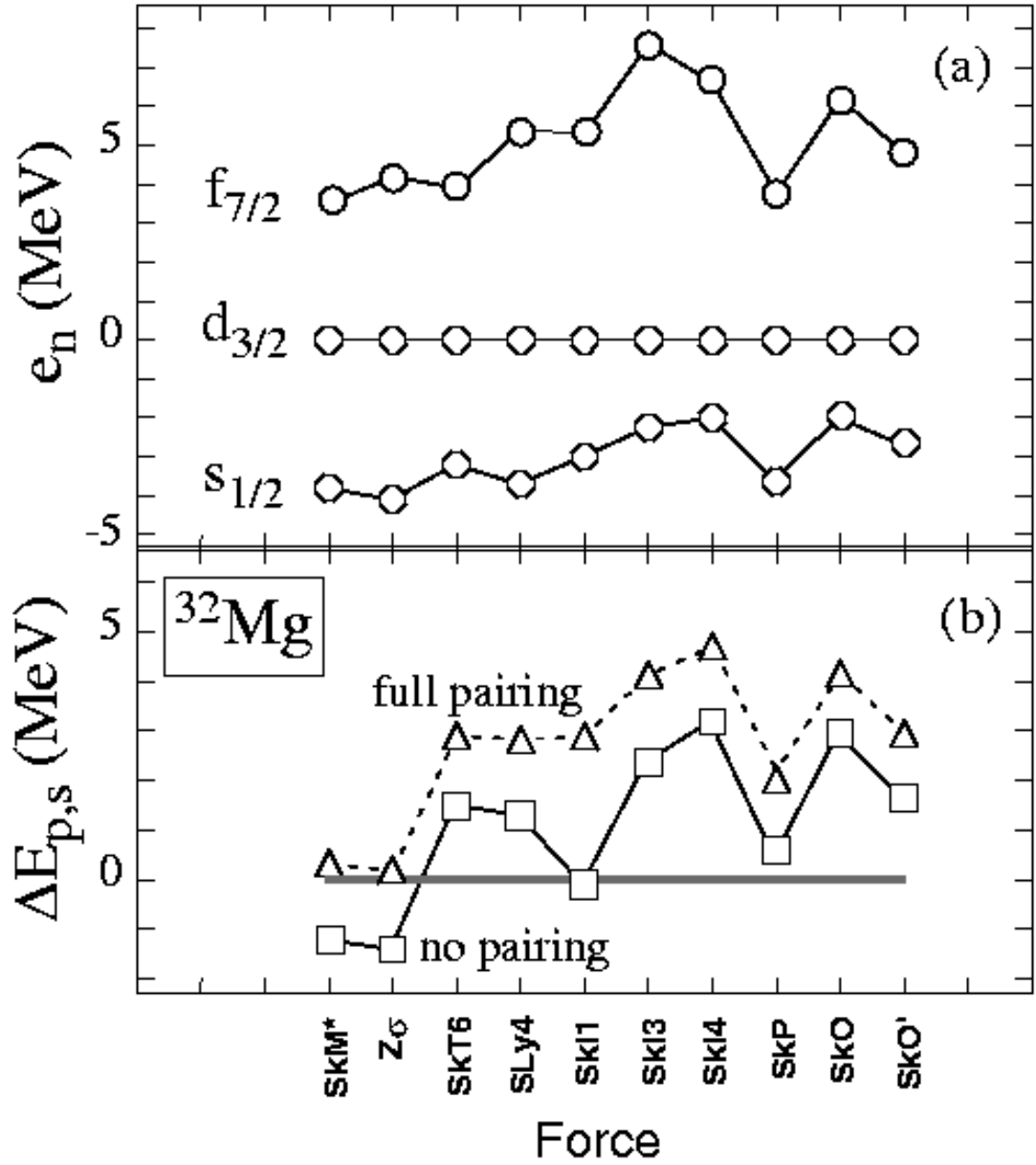}
\end{center}
\caption{Top: Spherical neutron shell structure in $^{32}$Mg calculated in
several Skyrme-HF models. The single-particle levels are normalized
to the energy of the $d_{3/2}$ shell. Bottom:  The prolate-spherical
energy difference calculated with (dashed line) and without (solid line)
pairing.
}
\label{shell-Mg}
\end{figure}

The spherical neutron shell structure for $^{32}$Mg predicted
by various Skyrme parameterizations
is shown in Fig.~\ref{shell-Mg}. Of particular interest is the size of the
$N$=20 magic gap which is measured by the distance between the
$f_{7/2}$ and $d_{3/2}$ shells:
\begin{equation}\label{gap20}
\Delta{e}_{20} \equiv e(f_{7/2}) - e(d_{3/2}).
\end{equation}
The variations of $\Delta{e}_{20}$ are nicely correlated
with the behavior of the height of the prolate minima
$\Delta E_{\text{prol}}$
in $^{32}$Mg, shown in  Fig.~\ref{shell-Mg} for two
variants of calculations: with  and without pairing (the
latter to single out
 the pure effect of the
particle-hole channel). Indeed, the
large values of
$\Delta E_{\text{prol}}$ in SkI3, SkI4, and SkO can be correlated with
large  values of  $\Delta{e}_{20}$. Likewise, small shell gaps in
SkM$^*$ and Z$_\sigma$ are consistent with
$\Delta E_{\text{prol}}$$\approx$0  obtained in
these models. However, there are exceptions to this rule. For instance, the
value of $\Delta{e}_{20}$ is rather low in SkT6 but the prolate minimum
is calculated to be at $\sim$3\,MeV.

In order to better understand some of the deviations between the
pattern of
 $\Delta{e}_{20}$ and $\Delta E_{\text{prol}}$,
it is instructive to return to
Fig.~\ref{asurf}.
The ``stiffest" interactions
are SkT6, SLy4, and SkP, and -- indeed --
for all these forces, spherical
ground states are predicted.
The ``softest" parameterizations are Z$_\sigma$, SkO, and SkO', but
the large value of  $\Delta{e}_{20}$ in SkO and SkO' gives rise to
spherical ground states.

\begin{figure}
\begin{center}
\leavevmode
\epsfxsize=16cm
\epsfbox{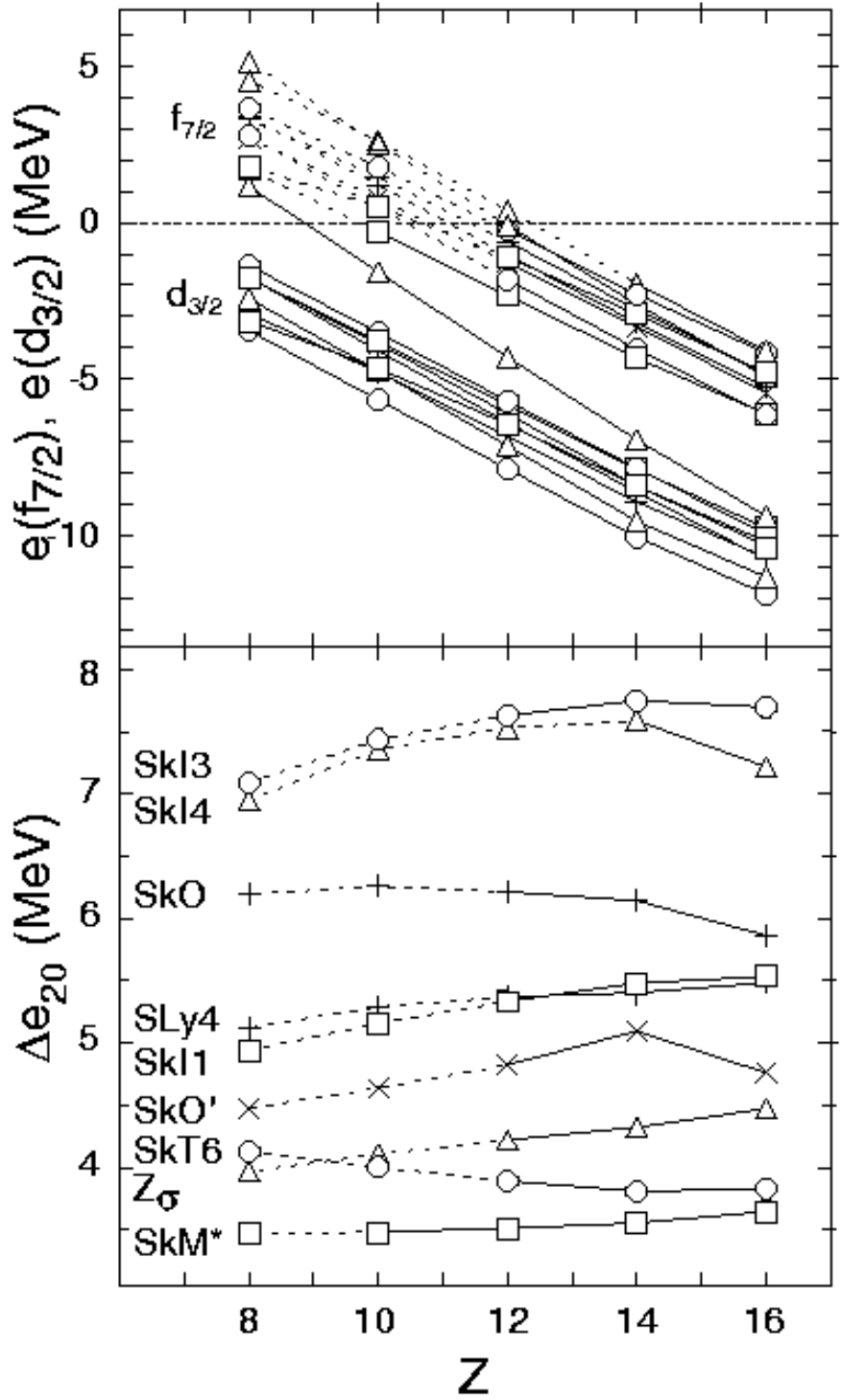}
\end{center}
\caption{Top: Single-neutron  $d_{3/2}$ and  $f_{7/2}$
energies as functions of $Z$  predicted in several
Skyrme-HF models. The positive-energy levels are marked by
a dashed line.
Bottom: the  size of the corresponding
single-particle $N$=20 gap, $\Delta e_{20}$.
}
\label{fig_espHF}
\end{figure}

The summary of single-neutron  $d_{3/2}$ and  $f_{7/2}$
energies 
for the $N$=20 isotones
calculated with several Skyrme forces is shown in
Fig.~\ref{fig_espHF}. As expected, the 
absolute binding energy
of these shells decreases rapidly when approaching
the drip-line nucleus $^{28}$O.  
For all the interactions considered, however, 
$\Delta{e}_{20}$ varies very slowly with $Z$.

\subsubsection{Deformation in the $N$$\approx$28 region}

\begin{figure}
\begin{center}
\leavevmode
\epsfxsize=16cm
\epsfbox{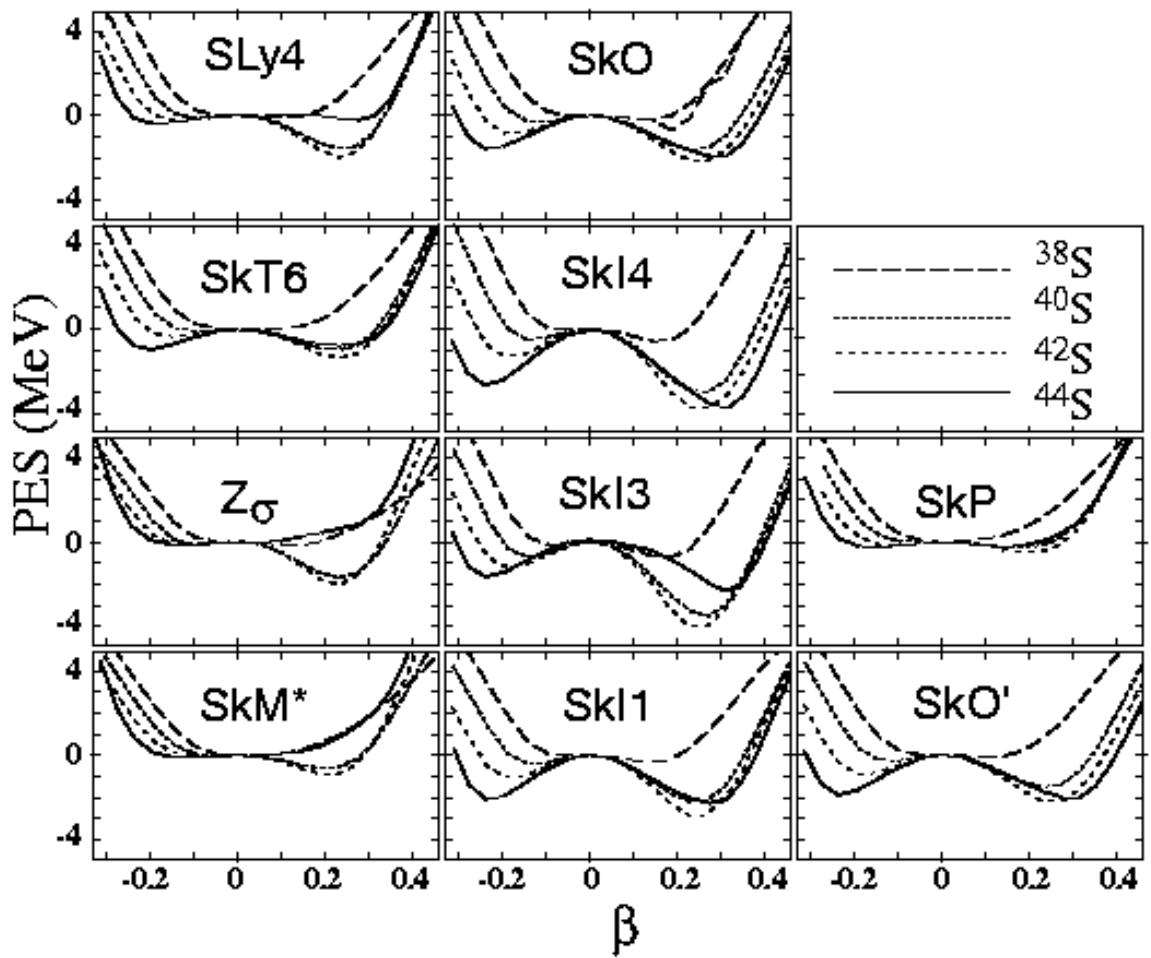}
\end{center}
\caption{Same as in Fig.~\protect\ref{PES-Mg}, except for
 $^{38,40,42,44}$S.
}
\label{PES-S}
\end{figure}

The results of calculations for $^{38,40,42,44}$S shown in
Fig.~\ref{PES-S} indicate that the $N$=28 shell gap is broken
around $^{44}$S. Indeed, most interactions
used  predict a deformed ground state
 for  $^{44}$S. It is worth noting that
 the two parameterizations that yield
strongest deformation effects in $^{32}$Mg, namely
SkM$^*$ and $Z_\sigma$,   do not produce
 deformed minima in $^{44}$S but rather
$\beta$-unstable PESs.
This indicates that
the deviations between results  should   be linked to
 the details of the underlying shell structure
which looks, of
course, different for the different shell closures.

\begin{figure}
\begin{center}
\leavevmode
\epsfxsize=16cm
\epsfbox{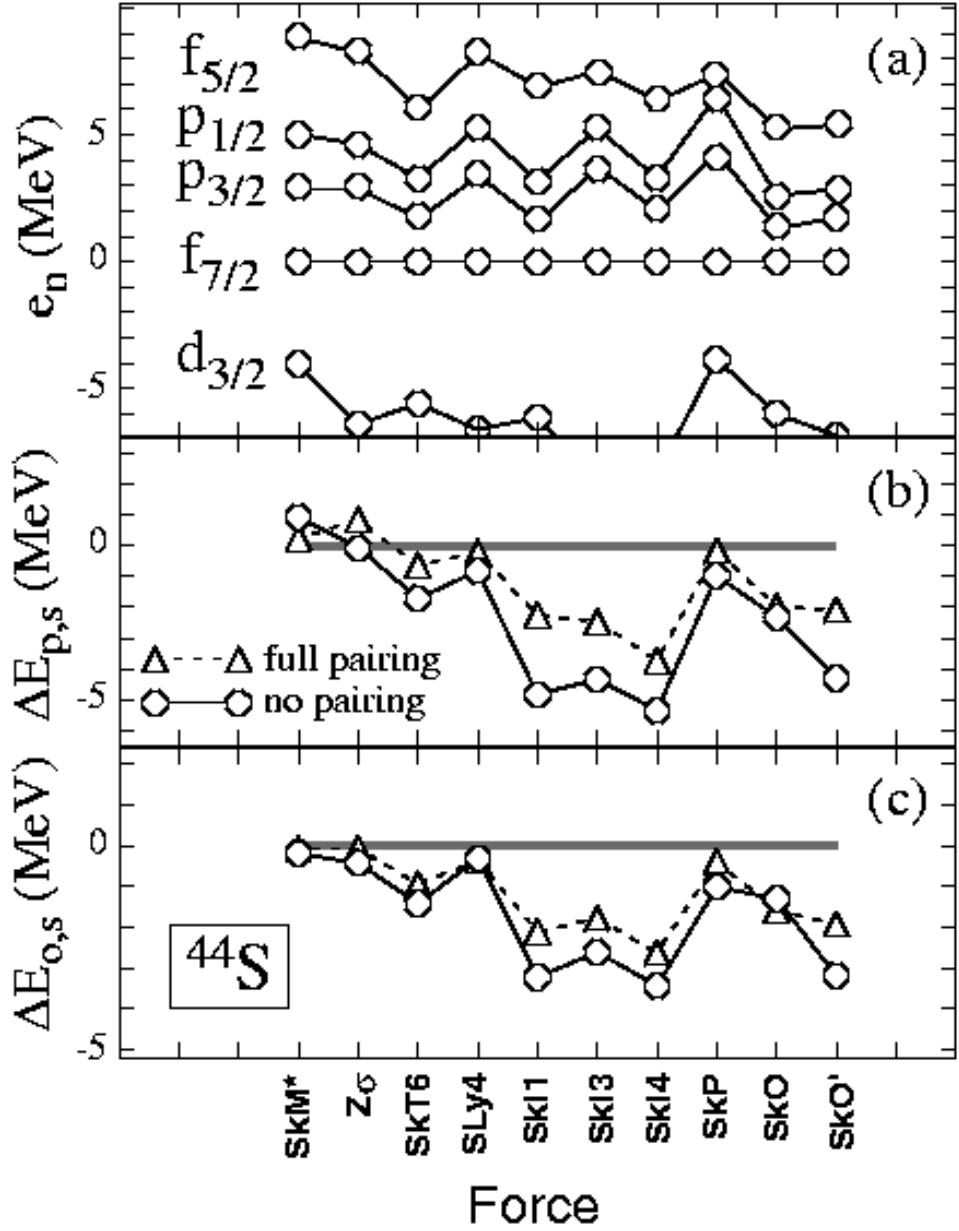}
\end{center}
\caption{Spherical neutron shell structure in $^{44}$S calculated
in several Skyrme-HF models.
 The single-particle levels are normalized
to the energy of the $f_{7/2}$ shell (a);
 the position of the deformed prolate
minimum with respect to the spherical HF state  (b);
the position of the deformed prolate
minimum with respect to the spherical HF state (c).
}
\label{shell-S}
\end{figure}

Figure~\ref{shell-S} shows the single-neutron structure in
$^{44}$S together with the calculated energies of the prolate,
$\Delta E_{\text{p,s}}$, and
oblate, $\Delta E_{\text{o,s}}$, minima (with respect to the spherical
configuration).
The position of the deformed minimum is greatly influenced
by the size of the $N$=28 gap \cite{[Wer94a]}:
\begin{equation}\label{gap28}
\Delta{e}_{28} \equiv e(p_{3/2}) - e(f_{7/2}).
\end{equation}
For most interactions considered,  $\Delta{e}_{28}$ is small -- of the order
of 2-3 MeV. Consequently, in most cases, the deformation energies
follow the pattern of $a_s$.

\subsubsection{The $N$$\approx$$Z$$\approx$40 region}

\begin{figure}
\begin{center}
\leavevmode
\epsfxsize=16cm
\epsfbox{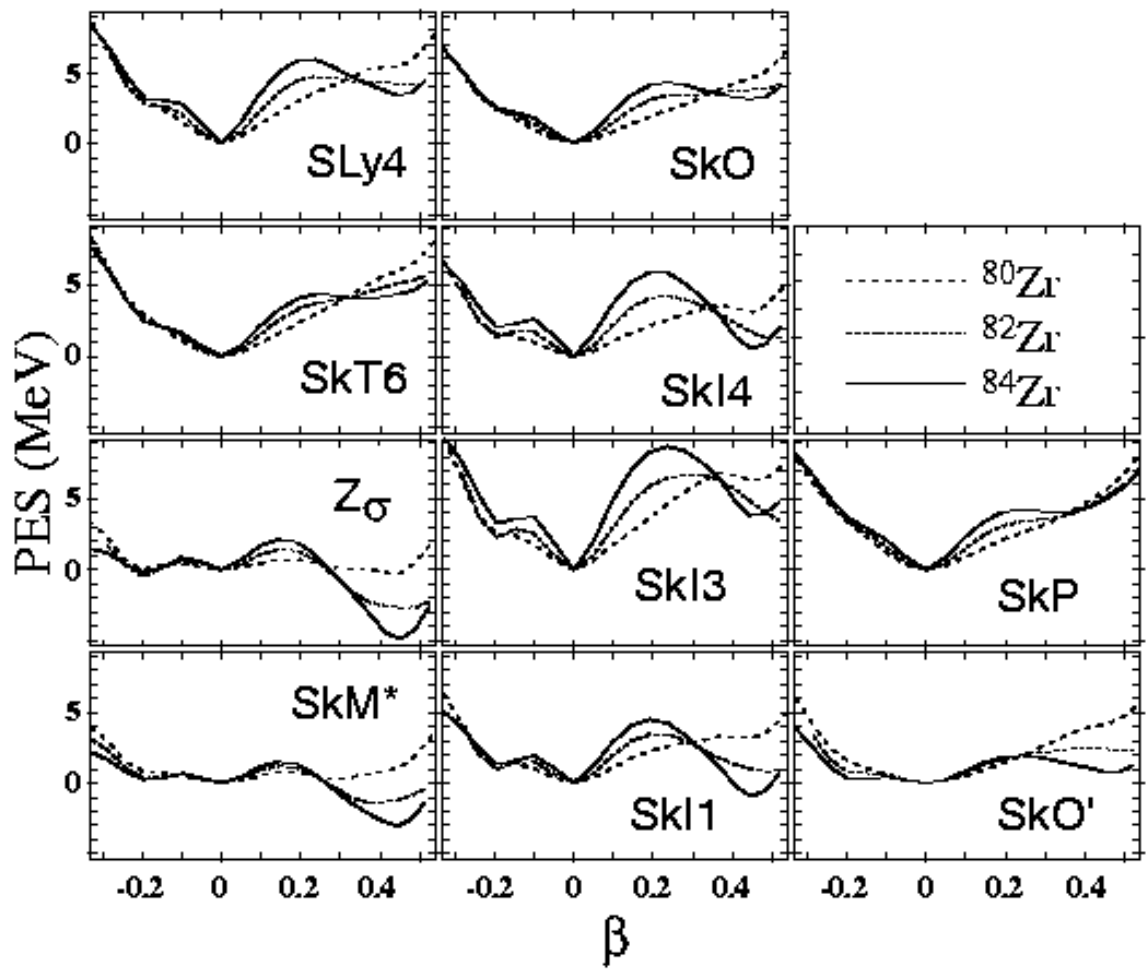}
\end{center}
\caption{Same as in Fig.~\protect\ref{PES-Mg}, except for
 $^{80,82,84}$Zr.
}
\label{PES-Zr80}
\end{figure}

The interplay between spherical and deformed subshell closures
at $N$ or $Z$=40 is  illustrated in Fig.~\ref{PES-Zr80}.
Although  coexisting
spherical and prolate minima in $^{80}$Zr are predicted for all the Skyrme
parameterizations used,
their relative position does  depend strongly on the interaction.
The interactions SkM$^*$, $Z_\sigma$, SkI1, SkI4, and SkO' predict
a strongly deformed ground state for $^{80}$Zr, in agreement with
experiment. Other forces, most notably SkP and SkT6,
 yield a spherical ground state.

\begin{figure}
\begin{center}
\leavevmode
\epsfxsize=16cm
\epsfbox{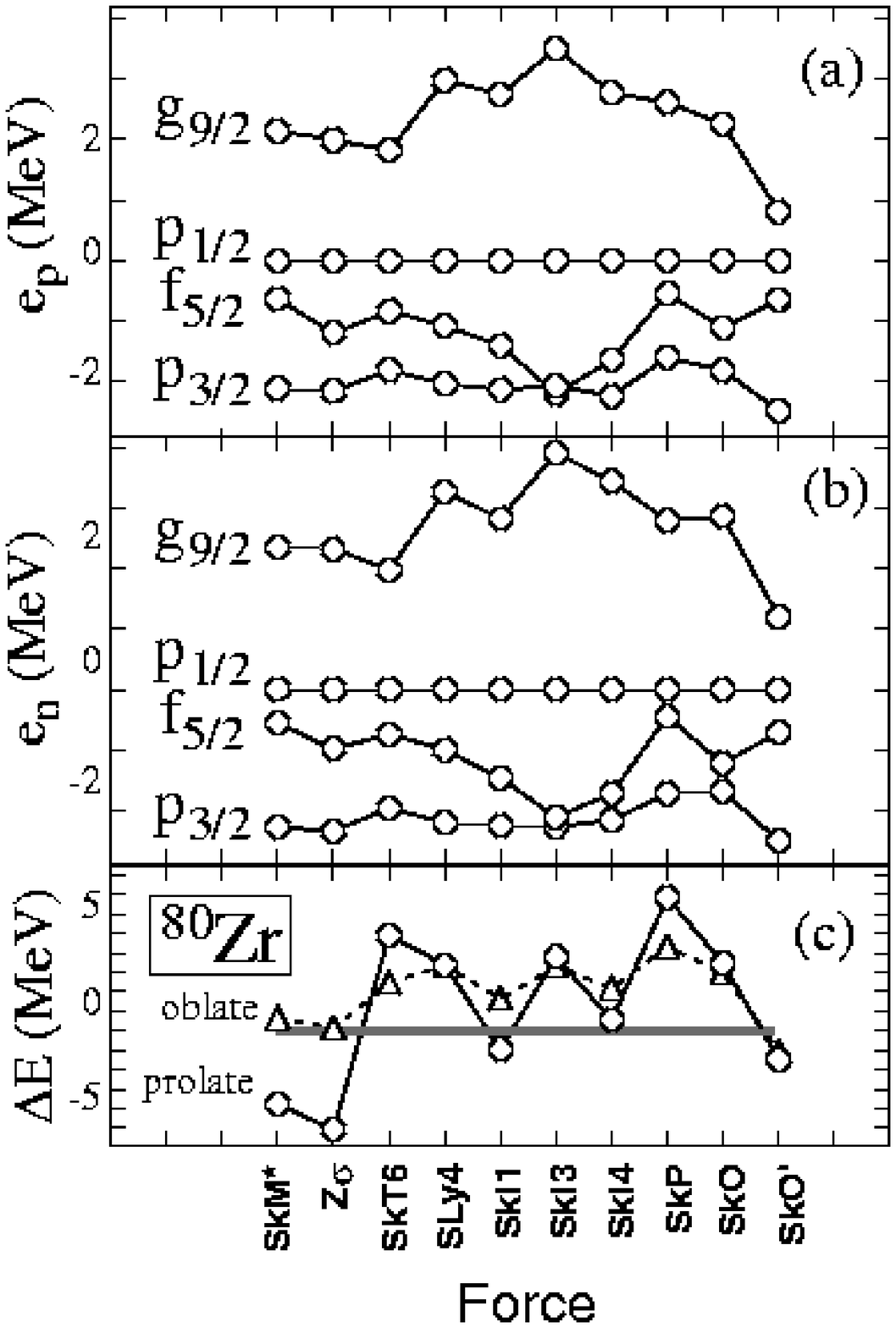}
\end{center}
\caption{Spherical neutron shell structure in $^{80}$Zr calculated
in several Skyrme-HF models.
 The single-particle levels are normalized
to the energy of the $p_{1/2}$ shell (a);
 spherical proton shell structure (b);
the position of the deformed prolate and oblate minima
 with respect to the spherical HF state (no pairing)  (c).
}
\label{shell-Zr80}
\end{figure}

The spherical shell structure in $^{80}$Zr is displayed in
Fig.~\ref{shell-Zr80}. Since for this nucleus $Z$=$N$, the proton
and neutron single-particle energies are very similar. (The influence
of Coulomb interaction on shell structure in this medium-mass
system is weak.) As in the case of $^{32}$Mg, there is a clear correlation
between the size of the $N$=$Z$=40 subshell closure,
\begin{equation}\label{gap40}
\Delta{e}_{40} \equiv e(g_{9/2}) - e(p_{1/2}),
\end{equation}
 the deformation energy, and
the surface-energy coefficients. For all Skyrme parameterizations which
predict a spherical ground state in $^{80}$Zr, either
$\Delta{e}_{40}$ is large (like in SkI3) or $a_s$ is large
(like in SkP), or both.

The PES and corresponding shell structure of $^{80}$Zr
provide a particularly clear example of how variations in the treatment
of the spin-orbit force can have a large impact on the
results. Compare the  ``twin'' parameterizations SkO and SkO'
 which
differ just by the switch $\theta_{\rm ls}$ in the spin-orbit
functional (\ref{eq:Eskyls}). The different spin-orbit force produces
a different splitting of the $1g$ levels, subsequently a different
shell gap at the Fermi surface (see  Fig.~\ref{shell-Zr80}), and
finally a different excitation energy of the prolate 
minimum (see
Fig.~\ref{shell-Zr80} and the PES in  Fig.~\ref{PES-Zr80}).

\subsubsection{The $N$$\approx$56, $Z$$\approx$40 region}

\begin{figure}
\begin{center}
\leavevmode
\epsfxsize=16cm
\epsfbox{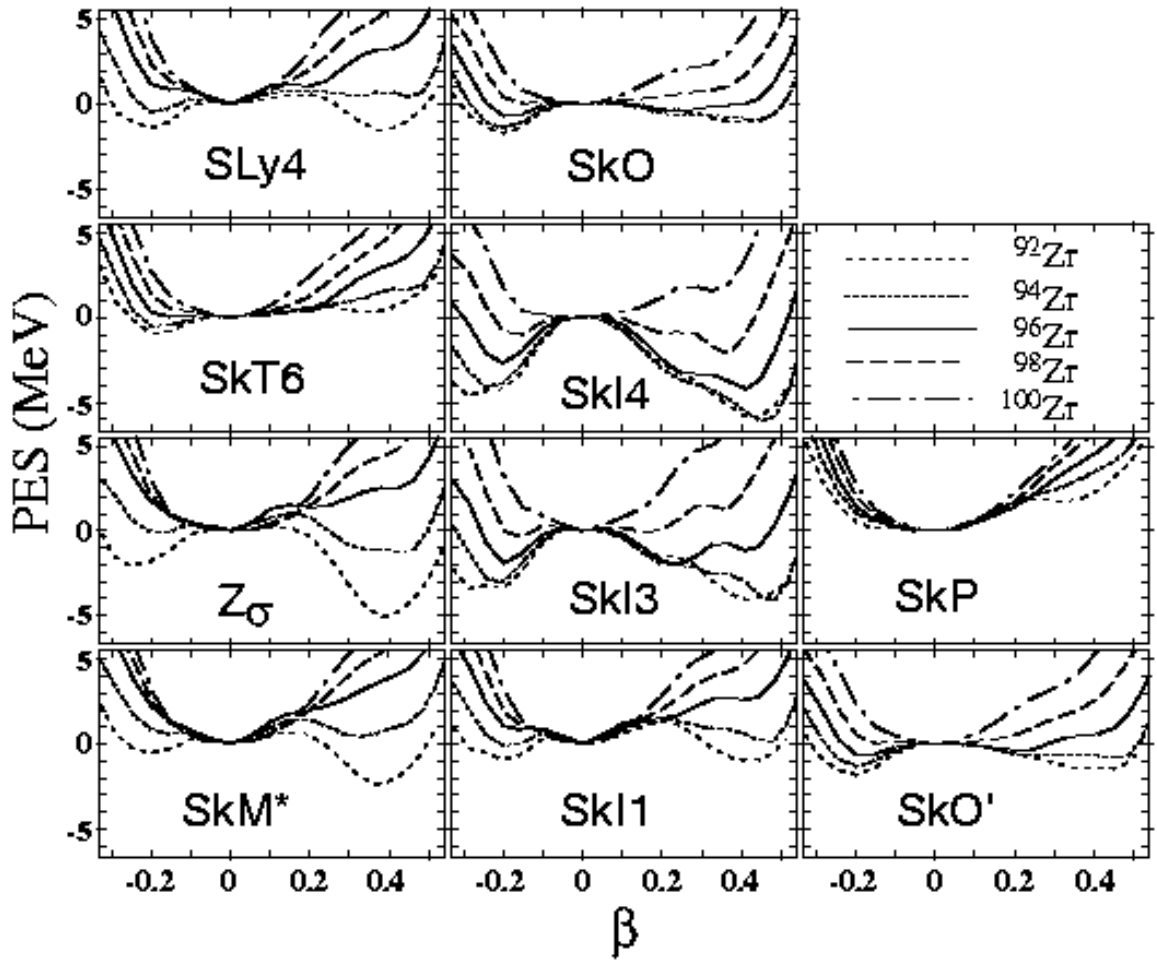}
\end{center}
\caption{Same as in Fig.~\protect\ref{PES-Mg}, except for
 $^{92,94,96,98,100}$Zr.
}
\label{PES-Zr96}
\end{figure}

In this region of shape coexistence, the best agreement with 
the observed
experimental trend
is given by SkM$^*$, $Z_\sigma$, SLy4, and SkI1 (see
Fig.~\ref{PES-Zr96}). Namely, $^{96}$Zr is predicted to be spherical,
 $^{100}$Zr very well deformed, and  $^{98}$Zr spherical, with 
a low-lying
deformed intruder state. The worst agreement with the data is
obtained in the SkP model in which
all isotopes considered have spherical ground states, and in the SkI4
model which predicts a strongly deformed ground state for
$^{94,96,98}$Zr.

Again, the general pattern of deformation energies can be explained
in terms of the calculated gap sizes: the $\Delta{e}_{40}$ proton gap and
the $N$=56 gap
\begin{equation}\label{gap56}
\Delta{e}_{56} \equiv e(s_{1/2}) - e(d_{5/2}).
\end{equation}
For instance,  for the  interaction SkI4
the proton $\Delta{e}_{40}$ and
the neutron $\Delta{e}_{56}$
are rather small and this yields a deformed
ground state in $^{96}$Zr. The opposite holds  for SLy4, which,
in addition, has a large value of $a_s$. Hence, it predicts spherical
$^{96}$Zr.

\subsection{Zero-point fluctuations}\label{sec:ZPE}

\begin{figure}
\begin{center}
\leavevmode
\epsfxsize=12cm
\epsfbox{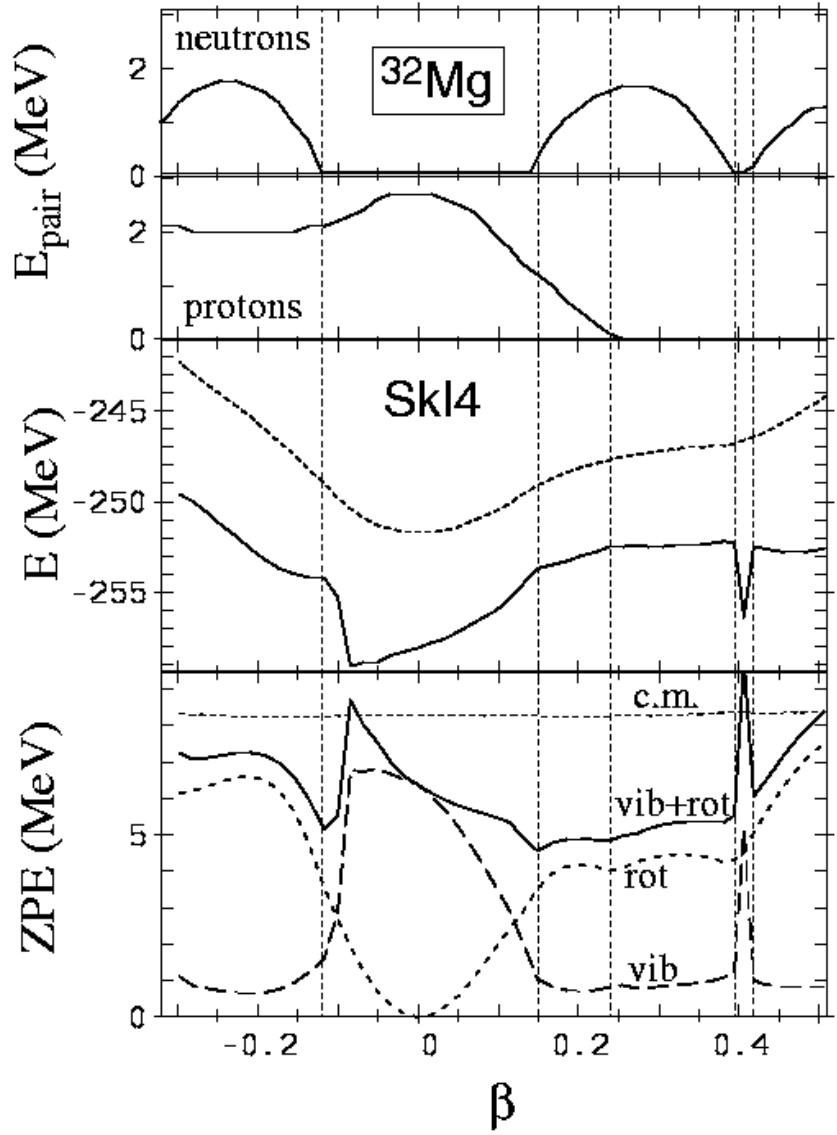}
\end{center}
\caption{The effect of ZPE corrections on the PES of $^{32}$Mg
calculated with the SkI4  SKyrme-HF model.
Top two: proton and neutron pairing energies.
Middle: uncorrected  (dashed line) and ZPE-corrected 
(solid line) PES.
Bottom: the various contributions to the ZPE
(rotational, dashed line; $\beta$-vibrational, long dashed line;
 center-of-mass, short dashed line) and the sum
of rotational and vibrational corrections, solid line. The vertical lines
mark borders of the regions where static pairing (in neutrons or protons)
vanishes.
}
\label{Mg-ZPE}
\end{figure}

\begin{figure}
\begin{center}
\leavevmode
\epsfxsize=12cm
\epsfbox{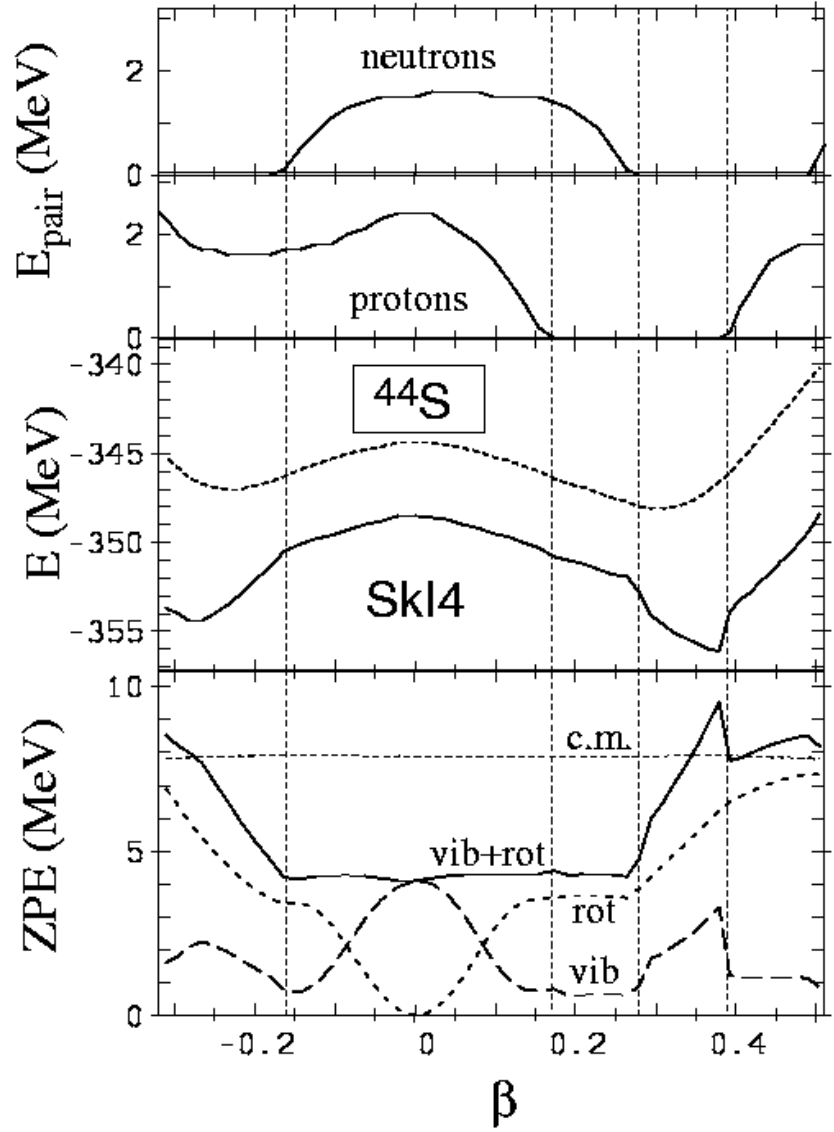}
\end{center}
\caption{Same as in Fig.~\protect\ref{Mg-ZPE}, except for
 $^{44}$S.
}
\label{S-ZPE}
\end{figure}

\begin{figure}
\begin{center}
\leavevmode
\epsfxsize=12cm
\epsfbox{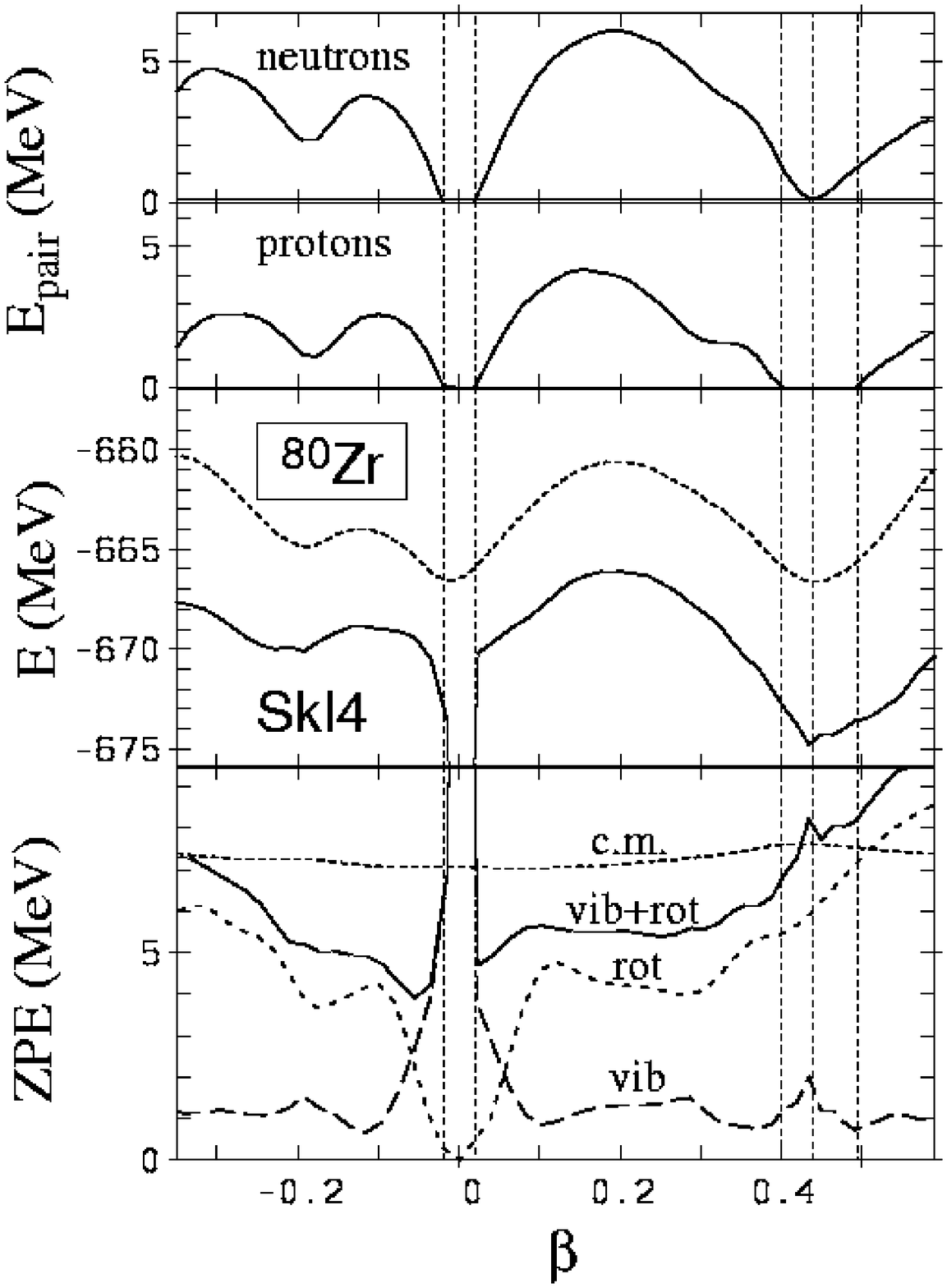}
\end{center}
\caption{Same as in Fig.~\protect\ref{Mg-ZPE}, except for
 $^{80}$Zr.
}
\label{Zr80-ZPE}
\end{figure}

\begin{figure}
\begin{center}
\leavevmode
\epsfxsize=12cm
\epsfbox{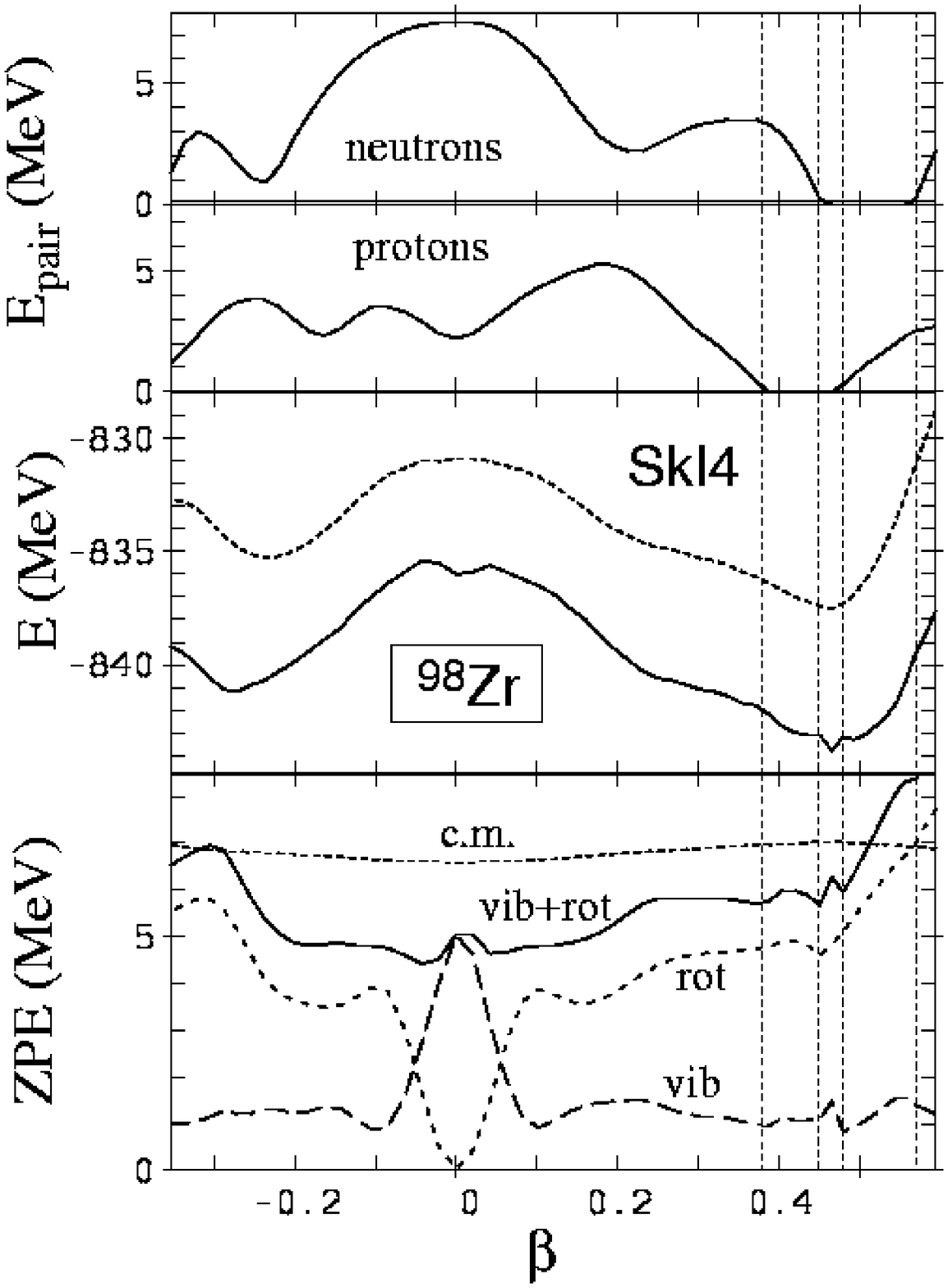}
\end{center}
\caption{Same as in Fig.~\protect\ref{Mg-ZPE}, except for
 $^{98}$Zr.
}
\label{Zr96-ZPE}
\end{figure}

The role of fluctuations beyond the mean field is illustrated
in Figs.~\ref{Mg-ZPE}-\ref{Zr96-ZPE} which show the effect of rotational,
$\beta$-vibrational, and center-of-mass corrections.
The calculations were performed with the SkI4 parameterization; a very
similar result (not shown here) was obtained with the SkM$^*$ force.

The  center-of-mass correction, Eq.~(\ref{eq:Ecm}), depends very weakly
on deformation; hence its contribution to the deformation energy can be safely
neglected. The rotational zero-point energy,
Eq.~(\ref{ZPErot}), is zero at the spherical shape and 
increases steadily with
deformation. The additional
fluctuations of $E_{\rm ZPE,rot}$ with $\beta$
are mainly due to the changes in the pairing field: the moment of inertia
$\Theta_{\rm rot}$, Eq.~(\ref{MOI}), increases when pairing correlations
are reduced, and this causes $E_{\rm ZPE,rot}$ to drop. The difference
of $E_{\rm ZPE,rot}$ between spherical and deformed minima is
around 4 MeV, i.e., this is a significant  correction to the total energy.
As discussed in Ref.~\cite{[Rei87]}, however, the rotational
zero-point energy should be supplemented by the vibrational counterpart
 $E_{{\rm ZPE},\beta}$, Eq.~(\ref{ZPEvib}). This quantity shows an opposite
behavior: it is strongly peaked around the spherical shape and  reaches
the value of $\sim$1\,MeV at large deformations. 
The large
peak at zero deformation  compensates 
for the correspondingly large dip
in the rotational ZPE such that, altogether, a smooth total
 ZPE emerges
whose main variation is the global trend 
to grow with  deformation.
The irregularities (kinks)
in $E_{{\rm ZPE},\beta}$, seen in Figs.~\ref{Mg-ZPE}-\ref{Zr96-ZPE},
are caused by the unphysical collapse of the BCS pairing in certain
regions of $\beta$, which, in turn, produces enormous spikes
in the collective quadrupole mass. 
Clearly, it is necessary to improve the description
of zero-point fluctuations by (i) taking into account  the particle-number
fluctuations, and (ii) by going beyond  the Inglis
cranking approximation.
Based on the present results, however, one can conclude that the
zero-point correction should be rather small for $^{32}$Mg and
$^{44}$S,
and that it 
favors the deformed
state by about 2\,MeV for 
 $^{80}$Zr and about 1\,MeV for  $^{98}$Zr.
The effect of shape fluctuations becomes more important
at large deformations due to the steady
 increase of $E_{\rm ZPE,rot}$. Consequently,
when studying superdeformations, fission barriers, fission valleys,
etc.,  zero-point corrections should be taken into account.

\section{Shell-Model analysis}\label{SM}

The mean-field analysis presented in the previous section is supplemented by
shell-model calculations for the neutron-rich nuclei
around $^{32}$Mg  using the
Shell Model Monte Carlo (SMMC)
technique \cite{[Lan93],[Koo97]}.
 In contrast to the mean-field approach, shell-model
calculations properly
treat  configuration mixing and dynamical fluctuations.
On the other hand, the rather small configuration space employed 
(here, two oscillator
shells) in comparison to the mean field
 can lead to an improper description of certain states.

\subsection{Shell-Model Monte Carlo Method}

The SMMC method
offers an alternative way to calculate
nuclear structure properties,
and is complementary to direct diagonalization.
SMMC cannot, nor is it designed to,
find  every energy eigenvalue of the
Hamiltonian. Instead, it is designed to give thermal or ground-state
expectation values for various
one- and two-body operators. Indeed, for larger nuclei, SMMC 
is presently
the only way to obtain information
on  properties of the system from a shell-model perspective.

The partition function of the imaginary-time
many-body propagator, $U=\exp(-\beta\hat{H})$,
is used to
calculate
the expectation values of any observable $\hat{\Omega}$:
\begin{equation}
\langle\hat{\Omega}\rangle=
{{\hbox{Tr}\hat{U}\hat{\Omega}}\over{\hbox{Tr}\hat{U}}}\;,
\end{equation}
where
\begin{equation}\label{HSM}
\hat{H} = \hat{H}_1 +  \hat{H}_2
\end{equation}
is the shell-model Hamiltonian containing one-body and two-body
terms,
and $T$=$1/\beta$ is the temperature of the system.
 The two-body term,
$\hat{H}_2$, is linearized through the Hubbard-Stratonovich
transformation, which introduces auxiliary fields over which
one must integrate to obtain physical answers. 
Since $\hat{H}$ contains many terms that do not
commute, one must discretize $\beta=N_t\Delta\beta$.
The method can be summarized as
\begin{eqnarray}
Z&=&\hbox{Tr}\hat{U}=\hbox{Tr}\exp(-\beta\hat{H})
\rightarrow
\hbox{Tr}
\left[\exp(-\Delta\beta\hat{H})\right]^{N_t} \nonumber \\
&\rightarrow&
\int{\cal D}[\sigma]G(\sigma)\hbox{Tr}\prod_{n=1}^{N_t}\exp
\left[\Delta\beta\hat{h}(\sigma_n)\right]\; ,
\end{eqnarray}
where $\sigma_n$ are the auxiliary fields. (There is one $\sigma$-field
for each two-body matrix-element in $\hat{H}_2$ when the two-body terms are
recast in quadratic form.)
${\cal D}[\sigma]$ is the measure of the integrand,
$G(\sigma)$ is a Gaussian in $\sigma$, and $\hat{h}$ is a one-body
Hamiltonian.  Thus, the shell-model problem is transformed
from the diagonalization of
a large matrix to one of large dimensional quadrature. Dimensions of the
integral can reach up to 5$\cdot$10$^4$ 
for the  $sdpf$ systems, and it is thus
natural to use Metropolis
random walk methods to sample the space. Such integration can most
efficiently be performed on massively parallel computers.
Further details are discussed in
Ref.~\cite{[Koo97]}.

The SMMC method is not free of extrapolation when realistic
Hamiltonians
are used.
The sign problem for realistic interactions was solved
by breaking the two-body
interaction into ``good'' (without a sign problem) and
``bad'' (with a sign problem) parts: $ H =  \hat{H}_{good} +
\hat{H}_{bad}$.  The  part $\hat{H}_{bad}$
 is  multiplied by a parameter, $g$,
with values typically lying in the range $-1 \le g \le 0$.
The  Hamiltonian $ \hat{H}(g) = f(g) \hat{H}_{good} + g  \hat{H}_{bad}$
has no sign problem for $g$ in this range. The function $f(g)$ is
used to help in extrapolations. It is constructed such that $f(g=1)=1$, and
takes the form $[1-(1-g)/\chi]$, with $\chi=4$
\cite{[Alh94],[Dea95]}.
The SMMC observables
are evaluated for a number of different negative $g$-values,  and the true
observables are obtained by extrapolation to $g$=1.
A prescription has been used to
remove center-of-mass contaminations inherent in the wave functions
when multi-$\hbar\omega$ spaces are used \cite{[Dea98]}. In each calculation
presented here, we took 6 values of $g$, and 4096 independent Monte Carlo
samples per $g$ value.

\subsection{The Effective Shell-Model Interaction}

In this work we wish to compare two shell-model interactions that
could prove useful for the $sdpf$ region. The first interaction was
derived using microscopic techniques \cite{[Dea98]}, while the second
is a more piece-wise interaction similar to those used in
highly truncated standard shell-model calculations
for nuclei near $N$=20.

Our first interaction, dubbed $sdpf$,
 is described in detail in Ref.~\cite{[Dea98]}.
In order to obtain a microscopic effective interaction, one
begins with a free nucleon-nucleon interaction $V$ which is appropriate
for nuclear physics at low and intermediate energies. The choice
made in Ref.~\cite{[Dea98]} was to work with the charge-dependent
version of the Bonn potential models as found in
 Ref.~\cite{[Mac96]}.
Standard perturbation techniques were then employed to obtain an effective
interaction in the full $sdpf$ model space. The interaction was
then modified in the monopole terms using techniques developed by
Zuker and co-workers \cite{[Zuk94],[Zuk98]}.

The second shell-model interaction employed in this work,
dubbed $sdpf'$, results from  a more standard, yet
less rigorous, approach to the problem.
Numerous shell-model studies have been carried out
in truncated model spaces for neutron-rich nuclei near
$N$=20 \cite{[War90a],[Fuk92],[Pov94]} and $N$=28
\cite{[Sch96],[Gla97],[Cau98]}.  Several
$sdpf$ shell effective interactions were used in these
studies;
 many of these interactions
 are quite similar in a number of respects.  All of
them use the Wildenthal USD interaction
\cite{[Wil83]} in the $sd$ part of the Hilbert space.  All
also use some `enhanced' version of the original
Kuo-Brown $pf$-shell $G$-matrix interaction \cite{[Kuo68]}
to describe nucleons in
that shell.  The cross-shell interaction is handled in one of two
different ways: matrix elements are generated  via a $G$-matrix
or via the Millener-Kurath potential \cite{[Mil75]}.
As is common in this type of
calculation, selected two-body matrix elements and single-particle
energies  have been further  adjusted to obtain agreement with
experiment.
Here, we use the following prescription: we incorporate the
USD interaction for the $sd$-shell \cite{[Wil83]}, and the
FPKB3 interaction as found in Ref.~\cite{[Pov81]}. We also used the
standard Millener-Kurath \cite{[Mil75]}
prescription for the cross-shell matrix elements.
However, our first investigations found that the scattering of particles
from the $sd$-shell to the $pf$-shell was  too strong. Therefore,
we reduced the cross-shell monopole matrix elements by 1.4 MeV.
 The single-particle energies
were adjusted to fit $^{41}$Ca single-particle energies

The $sdpf$ interaction describes satisfactorily the ground-state
masses in the $sd$-$pf$ region. The difference between theory
and experiement in the binding energies for the 10 nuclei
studied in Ref.~\cite{[Dea98]} is approximately $\pm 1.5$~MeV with a
statistical error of 0.75\,MeV. $B(E2)$ values were well
described across the $sd$-$pf$ region using standard effective
charges ($e_p$=1.5 and $e_n$=0.5). Occupation probabilities for
the $f_{7/2}$ shell were in fair agreement with highly truncated
interaction scenarios. The $sdpf'$ interaction cannot describe the
$B(E2)$ values across the $sd$-$pf$ region unless one invokes
two sets of effective charges ($e_p$=1.5, $e_n$=0.3
 in the $A$$<$40 region, and $e_p$=1.2, $e_n$=0.1
 in the $A$$>$40 region). 
Furthermore, binding energies were not well reproduced in the
$sdpf'$ interaction, although the excitation spectrum for a light
nucleus (e.g., $^{22}$Mg) was of the same quality as that of the
$sdpf$ interaction.  The occupation of the 
{\em full} $pf$-shell in the
neutron-rich nuclei is similar in both the $sdpf$ and $sdpf'$
interactions by construction, although more particles occupy
levels other than  $f_{7/2}$ in the $sdpf'$ case.
$B(E2)$ values and 
occupations numbers of three nuclei were used in the
fitting procedure of $sdpf'$: 
$^{36}$Ar, $^{32}$Mg, and $^{44}$Ti. Thus,
it is not surprising that the behavior of 
the two interactions is
similiar around $^{32}$Mg, while differences occur
 for other nuclei (see discussion below).

It should be clear that we prefer the $sdpf$ interaction as it is
based more on a theoretical derivation across the entire
 shell-model space in which the calculations were performed. However,
we believe it is worthwhile to investigate the differences
between this interaction and those obtained in a more
phenomenological way, such as $sdpf'$. We also note that
interactions derived in a similar fashion to $sdpf'$ have served
very useful purposes when calculations using them are performed
in truncated spaces
(e.g.,  as those by  Retamosa {\it et al.} 
\cite{[Ret97]}).
However, they are less able to reproduce
experimental data in full-space calculations such as those
performed here.

\subsection{Results of Shell-Model Calculations}\label{SMcalc}

\begin{figure}
\begin{center}
\leavevmode
\epsfxsize=14cm
\epsfbox{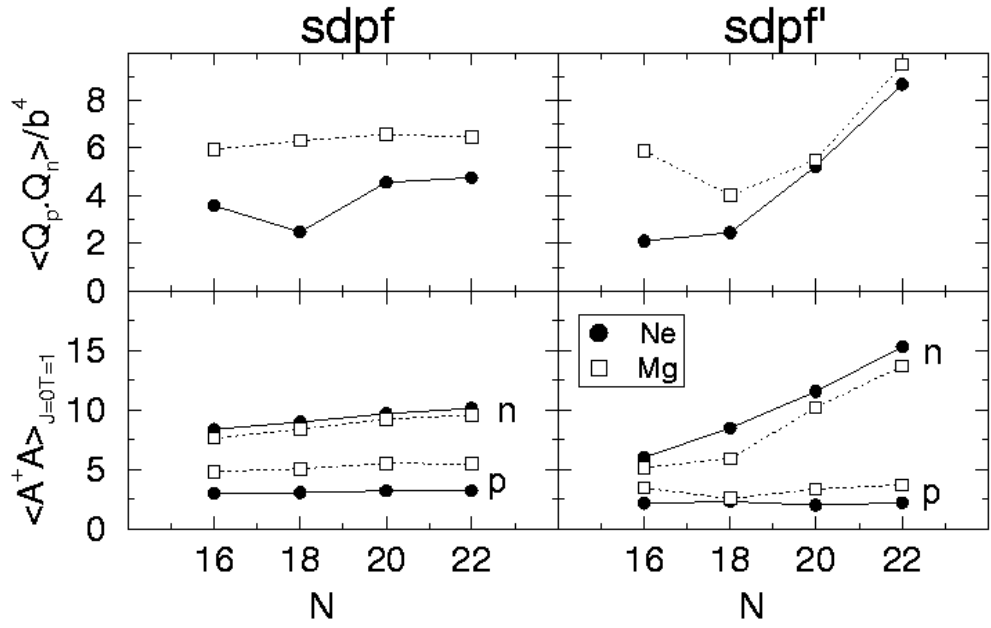}
\end{center}
\caption{Results of SMMC calculations with the $sdpf$ (left)
 and $sdpf'$ (right)
interactions for the neutron-rich Ne and Mg isotopes:
Top: $\langle Q_pQ_n\rangle$;
Bottom: $\langle A^+A\rangle$ ($J$=0, $T$=1).
}
\label{fig_SMres}
\end{figure}

\begin{figure}
\begin{center}
\leavevmode
\epsfxsize=14cm
\epsfbox{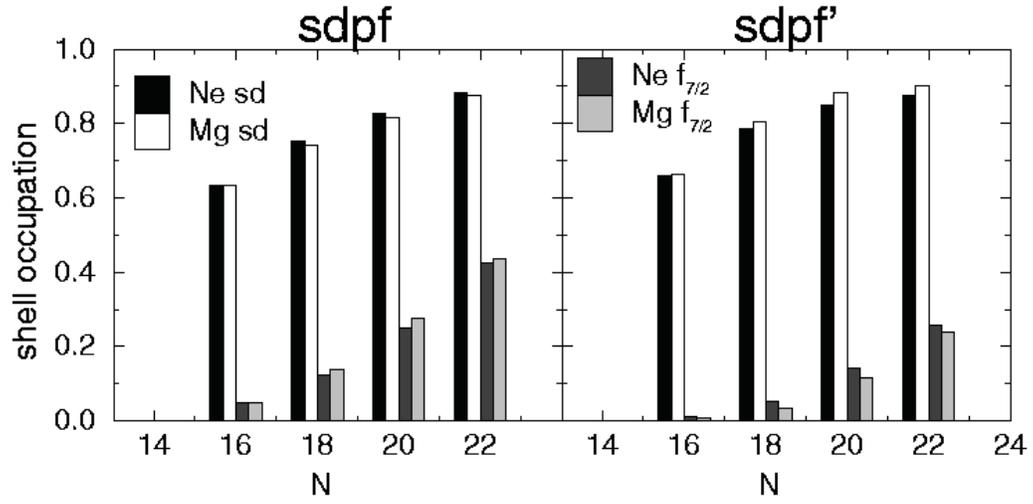}
\end{center}
\caption{Single-neutron shell-model occupations
in neutron-rich Ne and Mg isotopes calculated
in the SMMC with $sdpf$ (left) and
$sdpf'$ (right) effective interactions.
}
\label{fig_occ}
\end{figure}

The SMMC calculations were performed for a number of even-even nuclei
from the neutron-rich $N$=20 region. In order to relate the SMMC results to the
schematic shell-model scheme based on the broken-pair approach
\cite{[Hey87],[Hey91a]}, we show in Fig.~\ref{fig_SMres} the mean value of
$\hat{Q}_p\hat{Q}_n$, related to the proton-neutron quadrupole
interaction energy $E_Q$ of Eq.~(\ref{heyde}), and
the mean value of $\hat{A}^+_{01}\hat{A}_{01}$,
related to the pairing energy $E_{\rm pair}$ in the $J$=0, $T$=1 channel
($\hat{A}^+_{01}$ is the $J$=0, $T$=1 pair operator \cite{[Koo97]}). The calculations
were performed for the neutron-rich Ne and Mg isotopes.
The corresponding orbital occupation coefficients,
\begin{equation}
n_{j_\alpha} = \frac{N_{j_\alpha}}{2j_\alpha +1},
\end{equation}
where $N_{j_\alpha}$ is the average number of particles in the shell
$j_\alpha$, are displayed in Fig.~\ref{fig_occ}.

For the $sdpf'$ interaction, the result is consistent with the trend
predicted  by the schematic model. Namely, the  expectation value of
$\hat{Q}_p\hat{Q}_n$ increases at $N$=20 and 22, reflecting the increased
occupation of the $f_{7/2}$ shell.
For the $sdpf$ interaction, however,
the pattern is markedly different. In particular,
$\langle\hat{Q}_p\hat{Q}_n\rangle$
varies very little with $N$, especially for the Mg isotopes.
Although the $sdpf$ interaction predicts larger occupations of the $f_{7/2}$
shell, the value of $E_Q$ seems to be significantly greater
in the $sdpf'$ case. We shall come back to this apparent paradox in
Sec.~\ref{smmf}.

\begin{figure}
\begin{center}
\leavevmode
\epsfxsize=16cm
\epsfbox{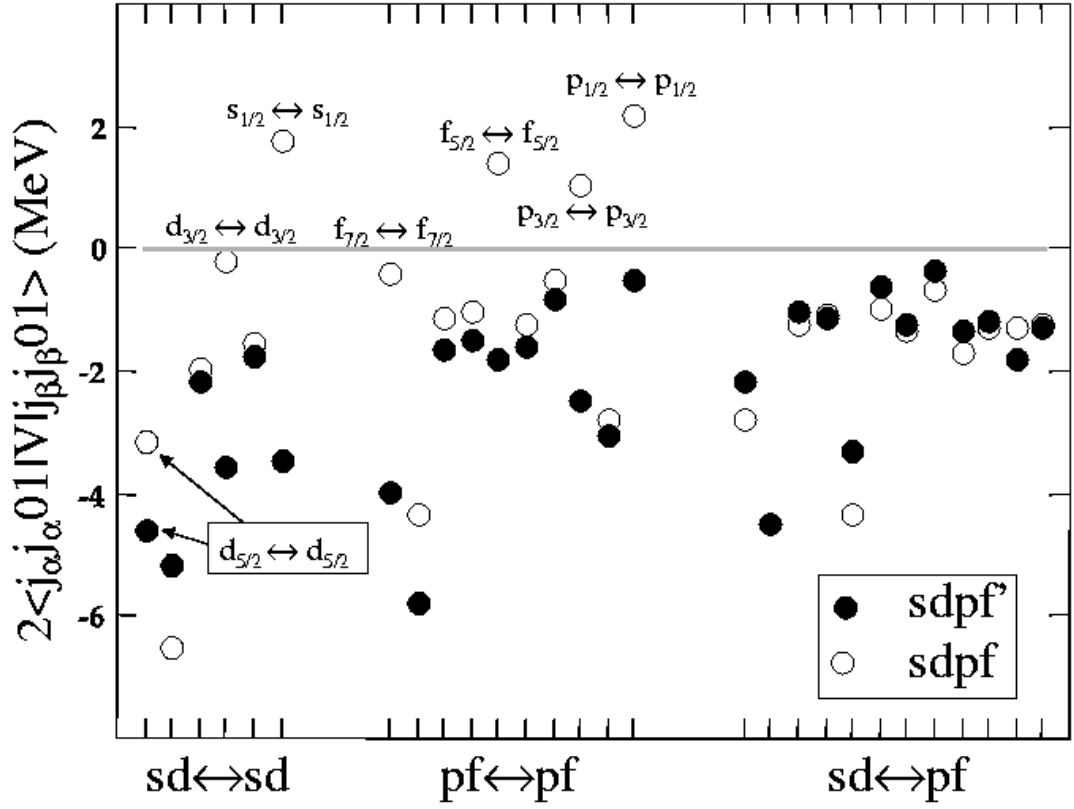}
\end{center}
\caption{$J$=0, $T$=1 matrix elements
$\langle j_\alpha j_\alpha |\hat{V}|j_\beta j_\beta\rangle$
of the  $sdpf$ (open circles) and $sdpf'$ (dots)
interactions. The diagonal matrix elements ($\alpha$=$\beta$)
are indicated. The matrix elements are presented according to the following
convention. Assuming that the single-particle orbitals are labeled as:
($d_{5/2}$, $d_{3/2}$, $s_{1/2}$,
$f_{f/2}$, $f_{5/2}$, $p_{3/2}$, $p_{1/2}$) $\equiv$
(1, 2, 3, 4 ,5, 6, 7),
the first six matrix elements are those within the $sd$ shell
[(1,1), (1,2), (1,3), (2,2), (2,3), (3,3)],
the next ten are the matrix elements within the $pf$ shell
[(4,4),(4,5)...,(7,7)], and the remaining twelve are the cross-shell
matrix elements [(1,4), (1,5),...,(3,7)].
}
\label{pairingme}
\end{figure}

Both interactions yield   fairly constant
$\langle\hat{A}^+_{01}\hat{A}_{01}\rangle$
 for the protons
(the proton pairing energy does not change with neutron number)
and an almost  linear increase with $N$ for the neutrons
(this behavior is indicative of a weak
neutron pairing). In order to understand an extremely weak dependence
of neutron $\langle\hat{A}^+_{01}\hat{A}_{01}\rangle$ predicted
in the $sdpf$ calculations, we show in Fig.~\ref{pairingme}
the $J$=0, $T$=1 matrix elements,
$\langle j_\alpha j_\alpha 01|\hat{V}|j_\beta j_\beta 01\rangle$,
of  $sdpf$ and $sdpf'$.
It is seen that, in general,
 the pairing interaction  within the $sd$ and $fp$ shells
 is weaker for $sdpf$, and the opposite is true for the
cross-shell  pair scattering. Moreover, except for the $d_{5/2}$ shell,
the diagonal pairing  matrix elements ($\alpha$=$\beta$)
of $sdpf$ are either close to zero or positive (i.e., the pairing
interaction in these states is actually  repulsive!).

\subsection{Mean-field analysis of shell-model
results}\label{smmf}

The shell-model Hamiltonian (\ref{HSM}) can be written as
\begin{equation}\label{H2}
\hat{H}=\sum_\alpha \epsilon_\alpha
 a^\dagger_\alpha a_\alpha +{1\over
4}\sum_{\alpha\beta\gamma\delta}
\bar{V}_{\alpha\beta\gamma\delta}\, a^\dagger_\alpha
a^\dagger_\beta a_\delta a_\gamma,
\end{equation}
where the single-particle indices (indicated by Greek letters)
denote the single-particle quantum numbers
($n,l,j,m,\tau=t_z$), $\epsilon_\alpha$ are the single-particle
shell-model
energies, and $\bar{V}_{\alpha\beta\gamma\delta}$ are the
(antisymmetrized) two-body matrix elements of the two-body
interaction. 

In order to translate shell-model results to the language
of mean-field theory, we carried out the HFB
calculations  using
the shell-model Hamiltonian (\ref{H2}).
In the following, this variant of calculations
will be referred
to as HFB-SM.
In the calculations
we impose 
spherical symmetry and disregard
neutron-proton pairing. The details of the HFB-SM
derivations are given in Appendix~\ref{AppB}.

\begin{figure}
\begin{center}
\leavevmode
\epsfxsize=16cm
\epsfbox{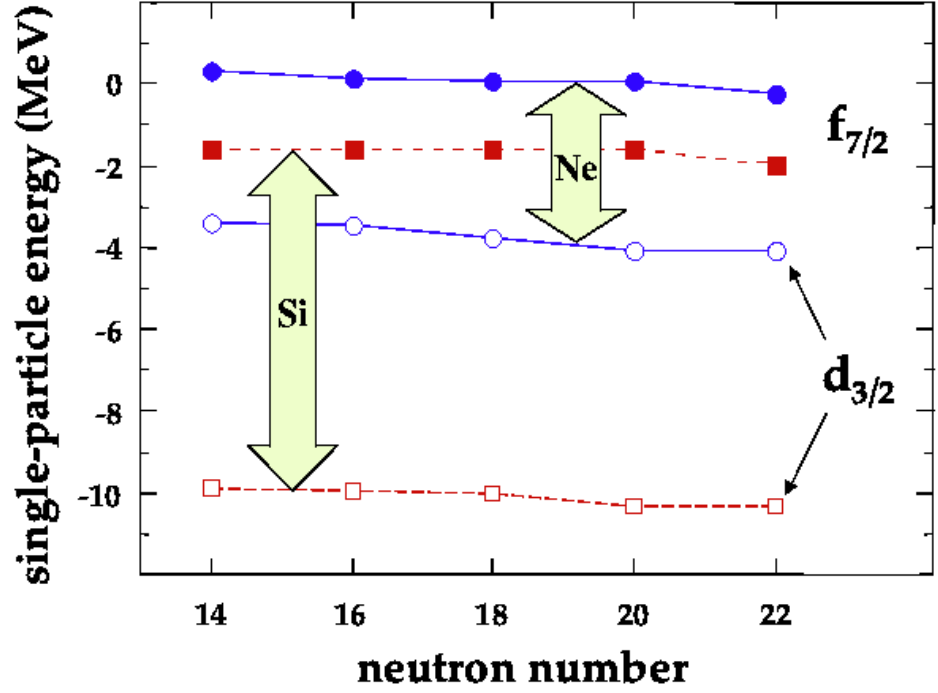}
\end{center}
\caption{Single-particle  $1d_{3/2}$ and  $1f_{7/2}$
neutron levels (\protect\ref{ecan}) predicted in  HFB-SM
with the sdpf' interaction
for the Ne and Si isotopes with 14$\le$$N$$\le$22. Note that
the size of the $N$=20 gap changes very little with $N$.
}
\label{fig_esp}
\end{figure}

\begin{figure}
\begin{center}
\leavevmode
\epsfxsize=16cm
\epsfbox{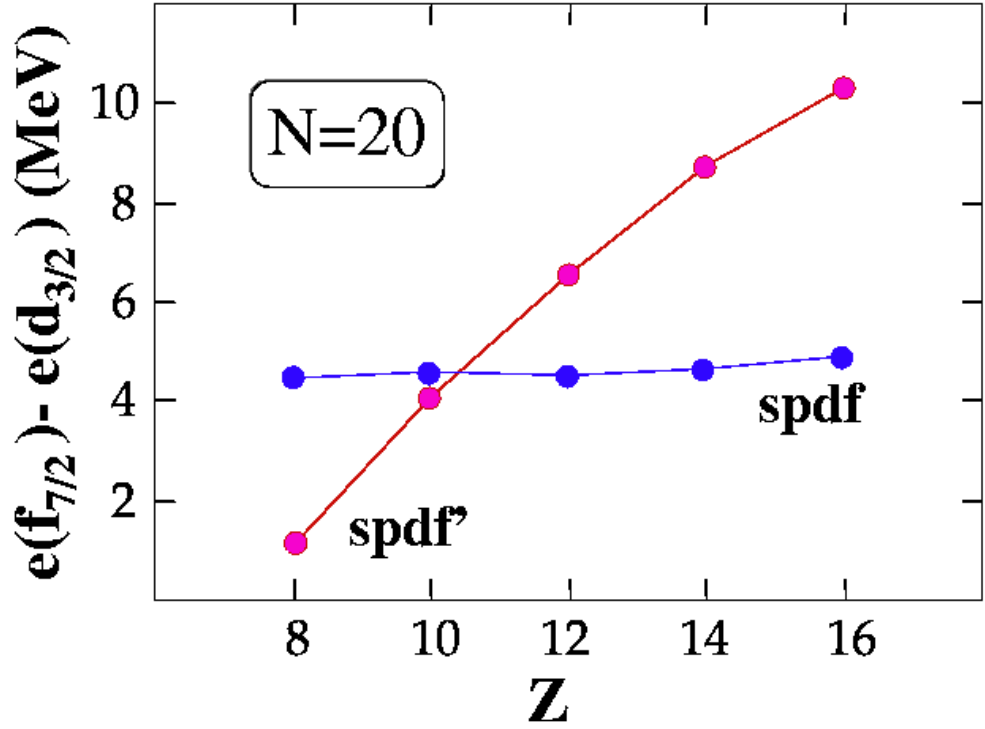}
\end{center}
\caption{Single-particle energy gap at $N$=20, $\Delta e_{20}$,
 calculated
in the HFB-SM with $sdpf$ and $sdpf'$ interactions for the $N$=20 isotones
of O, Ne, Mg, Si, and S.
Note that in $sdpf'$
the size of the $N$=20 gap depends dramatically on  $Z$.
}
\label{fig_N20}
\end{figure}

The canonical HFB single-particle  $d_{3/2}$ and  $f_{7/2}$ neutron
energy levels (\ref{ecan}) calculated in the $sdpf'$ variant
are shown in Fig.~\ref{fig_esp} for the Ne and Si isotopes.
Based on this result, two interesting conclusions can be drawn.
First, the isotonic dependence of single-particle levels
is very weak. Consequently, the size of the $N$=20 gap varies
little with $N$ (this conclusion also holds for the $sdpf$ interaction).
 Second, the single-particle energies strongly depend on
$Z$. This effect has been noticed in Ref.~\cite{[Hey91a]}, and was
discussed therein in terms of the monopole neutron-proton interaction,
that is, the shift in the spherical single-particle neutron energies due to
protons. It is seen that this monopole effect gives rise to the reduction
of the $N$=20 gap when decreasing $Z$.  Indeed, as shown in
Fig.~\ref{fig_N20}, the size of the $N$=20 neutron gap
calculated with the $sdpf'$ interaction  decreases from
$\sim$10\,MeV in $^{36}$S to  $\sim$2\,MeV in $^{28}$O.
It is important to emphasize that
this monopole effect predicted in HFB-SM,
 important for the excitation energy of the deformed
intruder configuration \cite{[Hey91a],[Woo92]},
is not a threshold phenomenon due to the weak binding;
the reduction of the magic gap comes solely from the shell-model
interaction.

\begin{figure}
\begin{center}
\leavevmode
\epsfxsize=16cm
\epsfbox{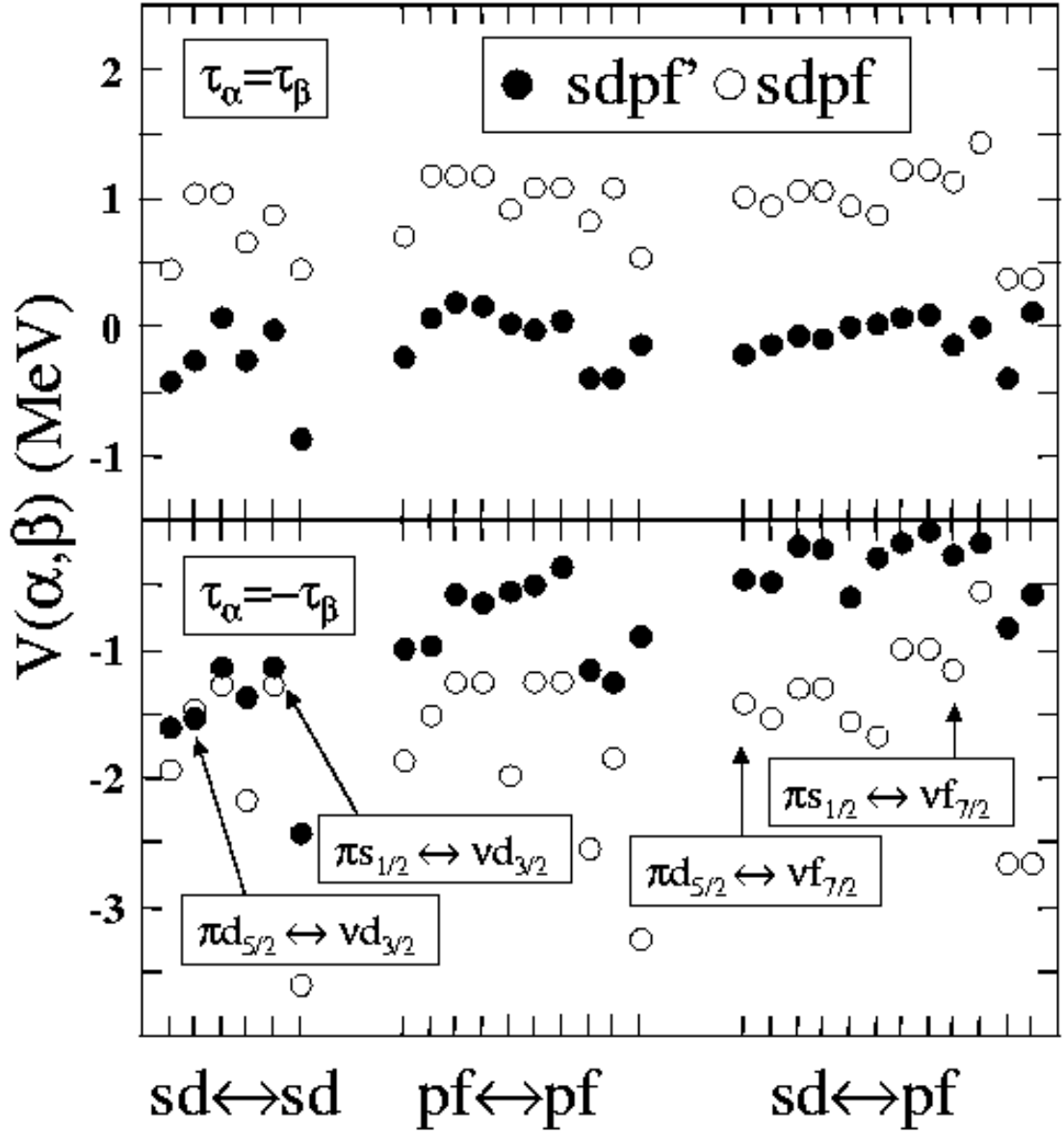}
\end{center}
\caption{Comparison between the particle-hole matrix elements $V(\alpha,\beta)$,
Eq.~(\protect\ref{veff}),
of $sdpf$ and $sdpf'$. The same convention
 is used as in
Fig.~\protect\ref{pairingme}. Top: neutron-neutron
(proton-proton) matrix elements ($\tau_\alpha$=$\tau_\beta$). Bottom:
proton-neutron  matrix elements ($\tau_\alpha$=--$\tau_\beta$).
}
\label{fig_forces}
\end{figure}

Figure~\ref{fig_N20} also shows the value of $\Delta e_{20}$
predicted with the $sdpf$ interaction. Here, the dependence of the gap
on the neutron number is very weak. To understand the difference between
predictions of the two interactions, Fig.~\ref{fig_forces}
shows the
matrix elements $V(\alpha,\beta)$ of
Eq.~(\protect\ref{veff}) for
$sdpf$ and $sdpf'$. These particle-hole matrix elements define
the self-consistent mean-field, hence the canonical single-particle energies.
Since the single-particle shell-model energies $\epsilon_\alpha$
do not vary with particle number, the variations of $e_\alpha$ with
$N$ and $Z$ are solely due to changes of the self-consistent mean-field.
In addition,
since the neutron-neutron contributions to $\Delta{e}_{20}$
do not depend on $Z$, the variation of the $N$=20 gap with  proton
number can be traced back to the proton-neutron interaction.
According to Eq.~(\ref{gamma}), the main contribution to the
$Z$-dependent part of  $\Delta{e}_{20}$ comes from the proton-neutron
terms:
\begin{equation}\label{vdiff}
V(\nu f_{7/2}, \pi j_\alpha) - V(\nu d_{3/2}, \pi j_\alpha).
\end{equation}
For nuclei discussed
in Fig.~\ref{fig_N20} the occupied proton shells
 are $d_{5/2}$ and $s_{1/2}$, and, precisely for these orbitals,
the difference (\ref{vdiff}) is close to zero for $sdpf$
and it is about 1\,MeV for  $sdpf'$. That is, it is very close to what
is seen in Fig.~\ref{fig_N20} ($\Delta{e}_{20}$ changes by
$\sim$1\,MeV/proton). One can thus conclude that the
monopole effect of Refs.~\cite{[Hey91a],[Woo92]} is very
weak for the $spdf$ interaction.

Coming back to the  prediction of the $sdpf$ interaction concerning
the unexpected behavior of $\langle\hat{Q}_p\hat{Q}_n\rangle$ versus $N$
(see Sec.~\ref{SMcalc}), it is instructive to inspect
the particle-hole matrix elements of
 Fig.~\ref{fig_forces}.
The proton-neutron
matrix elements of $sdpf'$ are negative (i.e., the particle-hole
interaction is attractive in this channel), and they are significantly
larger in magnitude than the like-particle matrix elements
(the latter ones are usually attractive or close to zero). This result
does not come as a surprise; it is generally believed that the
proton-neutron component of the particle-hole interaction
is dominant \cite{[Woo92],[Dob88],[Wer94]}. For $sdpf$, however,
the situation is  different: the proton-neutron interaction is,
generally, much stronger (especially in the $pf$ shell and
for the cross-shell matrix elements), but the particle-like
matrix elements are all {\em positive}.
Therefore, the structures predicted in  $sdpf$  result  from a subtle balance
between strongly attractive proton-neutron particle-hole interaction and
repulsive (and weaker, see Fig.~\ref{fig_forces}) proton-proton
and neutron-neutron particle-hole
forces. This is reflected in the SMMC results shown
in Table~\ref{tabQ}. In $sdpf'$, the
values of $\langle\hat{Q}_p\hat{Q}_n\rangle$
and $\langle\hat{Q}_n^2\rangle$ steadily increase when crossing the $N$=20 gap,
consistent with the increasing $f_{7/2}$ occupations. This is not the case
for $sdpf$ where the quadrupole collectivity decreases
in $^{34}$Mg in spite of the fact that
the  $f_{7/2}$ occupations are larger and the $\Delta{e}_{20}$ gap is
smaller than in the $sdpf'$ model.

\begin{figure}
\begin{center}
\leavevmode
\epsfxsize=16cm
\epsfbox{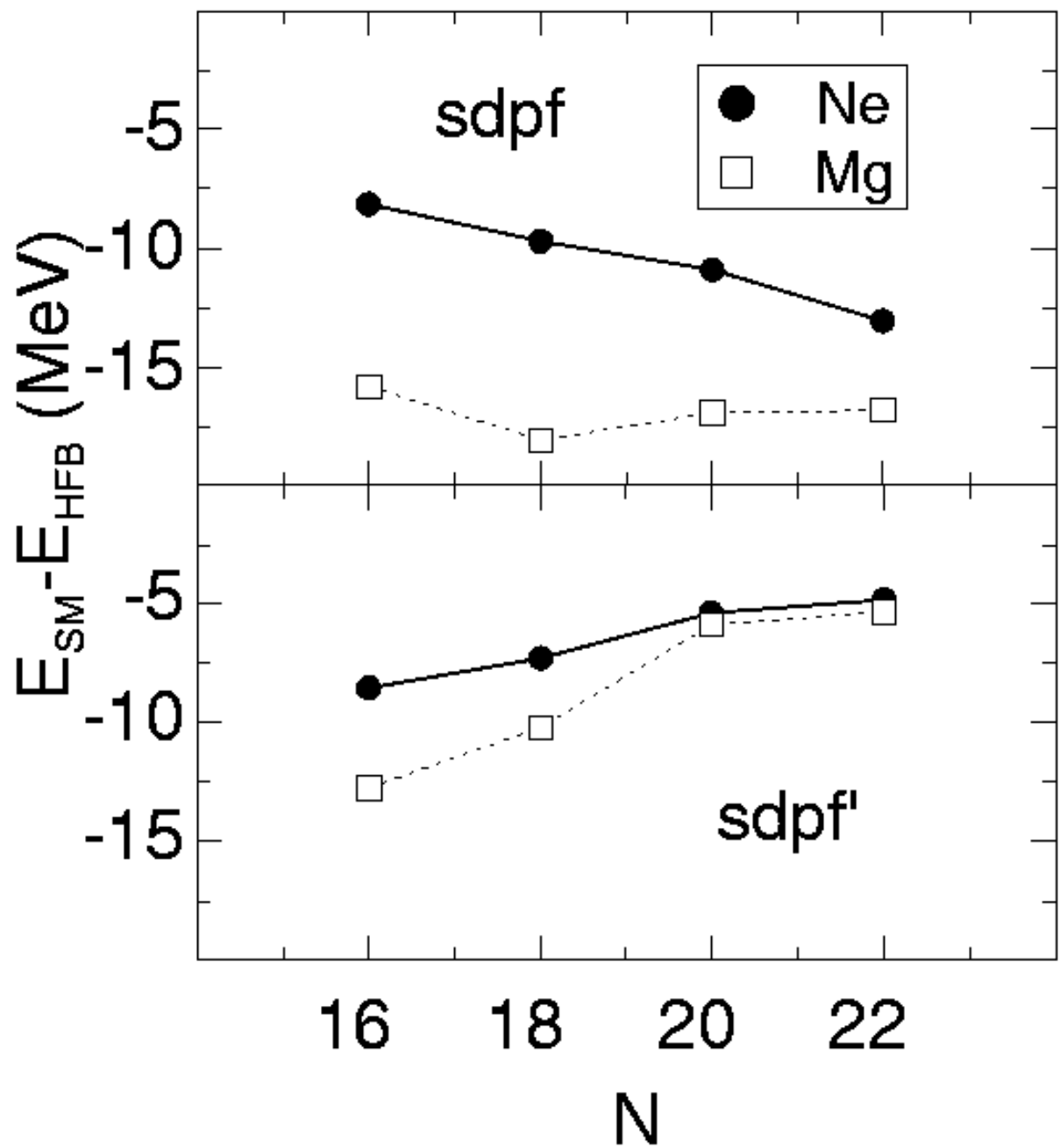}
\end{center}
\caption{Correlation energy  (\protect\ref{Ecorr})
 for the neutron-rich Ne and Mg isotopes. Top:
$sdpf$ results. Bottom  $sdpf'$ results.
}
\label{Fig_Ecorr}
\end{figure}

Figure~\ref{Fig_Ecorr} shows the SM correlation energy, i.e., the excess
of binding energy above  the spherical HFB estimate (\ref{EHFB}):
\begin{equation}\label{Ecorr}
E_{\rm corr} = E_{\rm SM} - E_{\rm HFB}.
\end{equation}
For $sdpf'$ the behavior of
$E_{\rm corr}$ does not follow the pattern of increased quadrupole collectivity
when crossing the $N$=20 gap. Actually, $E_{\rm corr}$ {\em decreases}. This
result is consistent with our mean-field results which predict the
coexistence of spherical and deformed shapes in $^{32,34}$Mg,
giving rise to the quadrupole-softness or shape mixing. Again, the behavior
of $E_{\rm corr}$ in $sdpf$ is  different. There is very little change
in  the SM correlation energy for the Mg isotopes; its rather large value
reflects the increased correlations due to the
significant occupation of the $f_{7/2}$ shell (spherical HFB-SM
calculations predict no
$pf$ neutrons in  $^{32}$Mg).
For both forces, the correlation energy in $^{32}$Mg is greater than
in $^{30}$Ne. This result corroborates our HF prediction that $^{30}$Ne
is  more spherical (i.e., coexistence effects are weaker).

To see the sensitivity of the SMMC predictions for $^{32}$Mg
 to the size of the splitting between the undisturbed single-particle
energies $\epsilon_\alpha$, we changed the splitting by $\pm$0.5 MeV around the
standard value. Surprisingly, such a variation changes
the $B(E2)$ (or $\langle Q^2\rangle$) value and the shell-model occupation
coefficients  very little. The correlation energy changes from
  --16.9\,MeV (standard
$N$=20 splitting) to  --16.6 (shell gap decreased by 0.5\,MeV) and
--12.4\,MeV  (shell gap increased by 0.5\,MeV). Hence, again,
in the $sdpf$ interaction the correlation energy is not
obviously related to the quadrupole collectivity.

\section{Conclusions}\label{conclusions}

In this paper, we have studied
the phenomenon of shape coexistence in semi-magic Mg-, S-, and
Zr-isotopes  employing two complementary theoretical approaches,
a self-consistent mean-field model (Skyrme-Hartree-Fock) and  
shell-model calculations which account for all correlations in a restricted
space.
The main conclusions of this study can be summarized as
follows.

The variety of Skyrme-HF predictions has been explored by
comparing all results for a set of 10 typical effective Skyrme
forces. For mean-field models, shape coexistence can be quantified in
terms of the relative energies of coexisting
local minima.  All selected Skyrme forces agree in
producing the same isotopic trends in these key features of shape
coexistence, but the actual preference for a spherical or deformed
ground state varies from force to force.  We have tried to relate the
results to other important features of the nucleus and find that
the main factor that determines the excitation energy of the
deformed intruder state in the HF calculations is the single-particle
shell structure (in particular, the sizes of the spherical
magic gaps and subshells). Another important quantity that defines the
nuclear deformability is the surface energy coefficient $a_s$.  Skyrme
interactions with large values of $a_s$ (SkT6, SLy4, SkP) favor
spherical configurations as compared to other forces (provided
that the corresponding shell effects are  similar). On the other hand,
 forces with  low values of  $a_s$ ($Z_\sigma$, SkO, SkO') give rise
to softer PES and low-lying intruder configurations.

The single-particle  structure can be  strongly affected by small
variations  in the definition of the energy functional. In
this context,  a good example
is the treatment of the spin-orbit term by
various parameterizations with respect to the inclusion of the $\bbox{J}^2$
contribution. For this purpose we had a twin pair of forces (SkO and
SkO') in the sample which differs just by this feature. It was
found that this modification  can have a large impact
on shape coexistence in some cases (here the most dramatic is $^{80}$Zr).

The proper treatment of pairing and zero-point correlations is crucial
if one aims at detailed predictions of shape coexistence. For instance,
according to our estimates,  the
zero-point rotational-vibrational
correction should be around 2\,MeV in $^{80}$Zr,
around   1\,MeV in
$^{98}$Zr, and  is expected to increase systematically with deformation.

\begin{figure}
\begin{center}
\leavevmode
\epsfxsize=16cm
\epsfbox{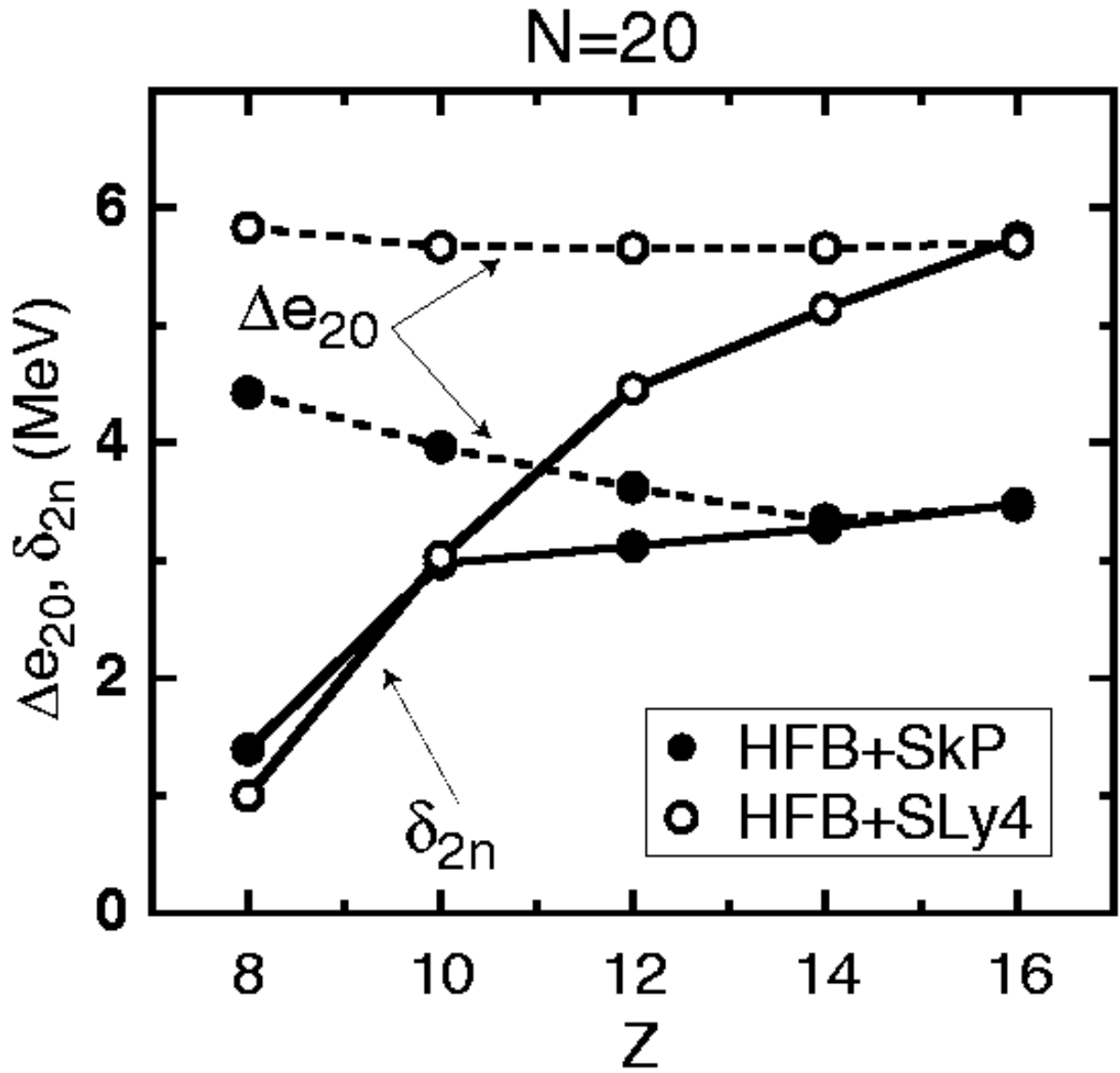}
\end{center}
\caption{Size of the $N$=20 gap, $\Delta e_{20}$, obtained
from canonical single-particle energies (dashed lines) and
the shell-gap parameter (\protect\ref{delta2n}) obtained from
 two-neutron separation energies (solid lines) calculated
in the HFB approach with Skyrme interactions
SkP (dots) and SLy4 (circles).
}
\label{N20HFB}
\end{figure}

For the 
  Skyrme interactions considered, the size of the $N$=20
gap varies very slowly with $Z$, and, except
for SkT6 and SkP,
 $\Delta e_{20}$ is quenched when approaching $^{28}$O
(see Figs.~\ref{fig_espHF} and \ref{N20HFB}).
This result agrees  with
the  $sdpf$ HFB-SM calculations. On the other hand,
 the size of the $N$=20 neutron gap
calculated with the $sdpf'$ interaction
decreases rapidly with $Z$.
This strong monopole effect can be traced back
to differences between certain proton-neutron matrix
elements of the shell-model interaction \cite{[Hey91a],[Woo92]}.
It is important to emphasize that this effect has its roots
in the properties of the shell-model Hamiltonian and  should
  not be confused \cite{[Cau98]} with the
 threshold phenomena  due to weak binding and the closeness of
the particle continuum.
Also, for a given isotopic chain, the $N$ dependence of the $N$=20 gap
has been found very weak for both shell-model interactions. This
contradicts recent conclusions of Ref.~\cite{[Cau98]} which
predict the sharp minimum of $\Delta e_{20}$ at $N$=20.
It should also be noted that the size of the single-particle gap
does not always correspond to the shell-gap parameter
 related to a difference between
two-neutron separation energies:
\begin{equation}\label{delta2n}
\delta_{2n} \equiv S_{2n}(N) - S_{2n}(N+2).
\end{equation}
Indeed, as seen in Fig.~\ref{N20HFB}, based on the spherical
Skyrme-HFB calculations, while  $\Delta e_{20}$ 
changes very weakly with $N$, $\delta_{2n}$ experiences
a dramatic drop when approaching $Z$=8. This indicates 
strong effects related to self-consistency in light drip-line nuclei.

The nucleus $^{32}$Mg has been found to be a classic
example  of shape coexistence;
the spherical and deformed configurations are close in energy and 
shape mixing is expected.
This prediction is consistent with the recent measurement
from GANIL \cite{Azaiez} according to which 
the $E_{4^+}/E_{2^+}$ 
ratio in  $^{32}$Mg falls well below the rotational limit.
A similar mixing effect is predicted to occur also
in $^{30}$Ne but is much weaker.
For most Skyrme parameterizations used,
the $N$=28 gap is predicted to be rather small. This gives rise to
strong deformation effects around $^{44}$S. The strong coexistence
effects are also predicted for $^{80}$Zr and $^{98}$Zr.

Both families of models applied in this work, i.e., self-consistent
mean-field models and the shell model, should be viewed as
{\em effective theories}. That is, their
predictive power crucially depends on the effective interaction assumed.
Since we do not know the ``true" energy functional (though we know
that it exists \cite{Hoh,Levy}), and we are still unable to derive
``exactly" the effective shell-model interaction and the
effective shell-model operators,
we are bound to try different parameterizations. From this point
of view, nuclear coexistence is a very challenging battleground. 
Although the global picture is understood,
 the structural details  strongly
depend on the actual phenomenology used and approximations
involved.

\acknowledgments

This research was supported in part by the U.S. Department of
Energy under Contract Nos.
DE-FG02-96ER40963
(University
of Tennessee), DE-FG05-87ER40361 (Joint Institute for Heavy
Ion Research), DE-AC05-96OR22464 with Lockheed Martin Energy
Research Corp. (Oak Ridge National Laboratory),
Bundesministerium f\"ur Bildung 
    und Forschung BMBF, Project No.\ 06 ER 808,
 the Polish Committee for Scientific
Research under Contract No.~2~P03B~040~14,
and by the NATO grant SA.5-2-05 (CRG.971541).

\appendix
\section{The Skyrme Parametrizations}
\label{sec:appforces}

For completeness, we provide  the parameters for
the sample of ten representative Skyrme forces
 used in this study.
The parameters $b_i$ and $b'_i$ used in the  definitions
of Sec.~\ref{Skyrme_model}
are chosen to give the  most compact formulation of the energy
 functional,
the corresponding mean-field Hamiltonian, and
 residual interaction. They
are related to the standard Skyrme parameters $t_i$ and $x_i$
\cite{[Que78],[Vau72],[Bar82],[Eng75]}
by:
\begin{equation}
  \begin{array}{rcl}
   b_0 & = &   t_0 (1+\frac{1}{2} x_0) , \\
   b_1 & = &  \frac{1}{4} \left[ t_1 (1+\frac{1}{2} x_1)+t_2 (1+\frac{1}{2} x_2)
              \right] , \\
   b_2 & = &  \frac{1}{8} \left[ 3t_1 (1+\frac{1}{2} x_1)-t_2 (1+\frac{1}{2} x_2)
              \right] , \\
   b_3 & = &  \frac{1}{4} t_3 (1+\frac{1}{2} x_3) , \\
   b_4 & = &  \frac{1}{2} t_4 ,               \\
&& \\
   b'_0& = &   t_0 (\frac{1}{2}+x_0) ,       \\
   b'_1& = &  \frac{1}{4} \left[ t_1 (\frac{1}{2}+x_1)-t_2 (\frac{1}{2}+x_2)
              \right]  ,                     \\
   b'_2& = &  \frac{1}{8} \left[ 3t_1 (\frac{1}{2}+x_1)+t_2 (\frac{1}{2}+x_2)
              \right] ,                      \\
   b'_3& = &  \frac{1}{4} t_3 (\frac{1}{2}+x_3) ,  \\
  \end{array}
\label{eq:bdef}
\end{equation}

Table \ref{tab:forces} displays 
 the parameters of the Skyrme
functional (\ref{eq:Esky})  given in the form
recoupled to the $t_i$, $x_i$ according to
Eq.~(\ref{eq:bdef}) (most of the existing codes
use this form of input).
All conventional Skyrme forces used simpler
pairing recipes. The pairing strengths $V_p$ and $V_n$ for the present
pairing treatment (see Sec.~\ref{sec:pairing}) have been adjusted 
anew to the neutron gaps in  $^{112,120,124}$Sn (using the values 
1.41, 1.39, and
1.31 MeV respectively)
and  proton gaps in $^{136}$Xe and $^{144}$Sm (using 0.98 and 1.25
MeV). The forces SkO and SkO' contained these gaps in the pool of
data throughout the  fit.

\section{The HFB approximation to the nuclear shell model}
\label{AppB}

The antisymmetrized two-body matrix element of the
shell-model Hamiltonian (\ref{H2}) can be written as:
\begin{eqnarray}
\bar{V}_{\alpha\beta\gamma\delta} & = &
\langle j_\alpha m_\alpha \tau_\alpha,
j_\beta m_\beta \tau_\beta
|\hat{H}_2|
j_\gamma m_\gamma \tau_\gamma,
j_\delta m_\delta \tau_\delta \rangle \nonumber \\
 &=& \sum_{JT} (-1)^{j_alpha+j_\gamma-j_\beta-j_\delta}(2J+1)
(2T+1)  \nonumber \\
&\times& \left( \begin{array}{ccc}
j_\alpha & j_\beta & J \\
m_\alpha & m_\beta & -M
\end{array}\right)
\left( \begin{array}{ccc}
j_\gamma & j_\delta & J \\
m_\gamma & m_\delta & -M
\end{array}\right) \nonumber \\
 & \times &
\left( \begin{array}{ccc}
{1\over 2} & {1\over 2}  & T \\
\tau_\alpha & \tau_\beta & -M_T
\end{array}\right)
\left( \begin{array}{ccc}
{1\over 2} & {1\over 2}  & T \\
\tau_\gamma & \tau_\delta & -M_T
\end{array}\right)  \nonumber \\
 & \times & \! \langle j_\alpha j_\beta JT|\hat{H}_2|j_\gamma j_\delta
JT\rangle
\sqrt{(1+\delta_{\alpha\beta})(1+\delta_{\gamma\delta})},
\end{eqnarray}
where the condition
\begin{eqnarray}
\langle j_\alpha j_\beta JT|\hat{H}_2|j_\gamma j_\delta
JT\rangle &=&
(-1)^{J+T-{j_\delta}-{j_\gamma}} \nonumber \\
&\times& \langle j_\alpha j_\beta JT|\hat{H}_2|j_\delta j_\gamma
JT\rangle
\end{eqnarray}
guarantees the antisymmetrization of matrix elements.

 Because of the  condition of sphericity,
and the fact that in the shell-model space considered each
spherical shell has a unique value of ($l$,$j$), the HFB
procedure is particularly simple. Namely, the quasiparticle
canonical states
are given by a  BCS transformation
\begin{equation}
c^\dagger_\alpha = u_\alpha a^\dagger_\alpha - v_\alpha a_\alpha.
\end{equation}
The amplitudes $(u_\alpha, v_\alpha)$ define the
self-consistent mean field:
\begin{equation}\label{gamma}
\Gamma_\alpha  =
\sum_{\beta}
    \bar{V}_{\alpha\beta\alpha\beta}\,v^2_\beta,
\end{equation}
the self-consistent pairing gaps
\begin{eqnarray}\label{gaps}
\Delta_\alpha &=&
\sum_{\beta}
    \bar{V}_{\alpha\bar{\alpha}\beta\bar{\beta}}\,u_\beta v_\beta \nonumber \\
 &=& -{1\over\sqrt{2j_\alpha+1}}
\sum_{j_\beta,\tau_\beta}
\delta_{\tau_\alpha,\tau_\beta} (-1)^{l_\alpha-l_\beta}
u_\beta v_\beta  \nonumber \\
&\times& \sqrt{2j_\beta+1}
\langle j_\alpha j_\alpha 01|H_2|j_\beta j_\beta 01\rangle,
\end{eqnarray}
and the total HFB energy:
\begin{equation}\label{EHFB}
E_{\rm HFB}=\sum_{\alpha}\left[\left(\epsilon_\alpha
+ {1\over 2}\Gamma_\alpha\right)
v^2_\alpha -{1\over 2} \Delta_\alpha u_\alpha v_\alpha\right].
\end{equation}

In deriving Eq.~(\ref{gaps}) we employed the phase convention of Condon-Shortley
for time reversal:
\begin{equation}
\hat{T}|nljm\rangle = (-1)^{l+j+m}|nlj-m\rangle.
\end{equation}
The particle-hole matrix element
 $\bar{V}_{\alpha\beta\alpha\beta}$
in Eq.~(\ref{gamma}) can be written as
\begin{equation}
\bar{V}_{\alpha\beta\alpha\beta}=(2j_\beta+1)V(\alpha,\beta),
\end{equation}
where
\begin{eqnarray}\label{veff}
 V(\alpha,\beta) &=&
  {{1+\delta_{\alpha\beta}}\over (2j_\alpha +1)(2j_\beta+1)} \\  \nonumber
 &\times& \left\{\delta_{\tau_\alpha,
\tau_\beta}\overline{V_{j_\alpha j_\beta}^{T=1}} +
 \delta_{\tau_\alpha,
-\tau_\beta}
{1\over 2}\left(\overline{V_{j_\alpha j_\beta}^{T=0}}
+ \overline{V_{j_\alpha j_\beta}^{T=1}}\right)
\right\},
\end{eqnarray}
and the $(2J+1)$-averaged  matrix elements   are
\begin{mathletters}\begin{eqnarray}
\overline{V_{j_\alpha j_\beta}^{T=1}} & = & \sum_J (2J+1)
\langle j_\alpha j_\beta J1|\hat{H}_2|j_\alpha j_\beta J1\rangle,
 \\
\overline{V_{j_\alpha j_\beta}^{T=0}} & = & \sum_J (2J+1)
\langle j_\alpha j_\beta J0|\hat{H}_2|j_\alpha j_\beta J0\rangle.
\end{eqnarray}\end{mathletters}
The neutron and proton Fermi levels,
$\lambda_{t_z}$, are determined from
the particle number equations
\begin{equation}\label{Pnum}
\sum_{\alpha}v_{\alpha,t_z}^2 = N_{t_z},
\end{equation}
where $N_{1/2}$ and $N_{-1/2}$ are the numbers of
valence neutrons and protons, respectively. The
HFB equations are reduced to a set of coupled
equations for occupation amplitudes:
\begin{equation}\label{BCS}
v^2_\alpha = {1\over 2}\left[
1-\frac{e_\alpha -
\lambda_{t_z}}{\sqrt{(e_\alpha-\lambda_{t_z})^2 +
\Delta_\alpha^2}}
\right],
\end{equation}
where
\begin{equation}\label{ecan}
e_\alpha = \epsilon_\alpha + \Gamma_\alpha
\end{equation}
are canonical single-particle energies.
Equations (\ref{Pnum}) and (\ref{BCS})
have been  solved iteratively.


\clearpage
\widetext
\begin{table}
\caption{\label{tab:forces}
Parameters of the Skyrme forces used in this study given in terms
of the functional as specified in Sec.~\protect\ref{Skyrme_model}
and \protect\ref{sec:pairing}. The column ``source'' lists the citations
where the parameterizations were first  defined.
}
\begin{tabular}{|l|dddd|dddd|}
  Force & $t_0$ & $t_1$  & $t_2$  & $t_3$ &  $x_0$ & $x_1$  & $x_2$  & $x_3$ \\
\hline
SkM* & -2645.0   & 410.0   & -135.0   & 15595.0  &
          0.090   &  0.0     &  0.0     & 0.0     \\
Z$_\sigma$ & -1983.76  & 362.25  & -104.27  & 11861.4  &
          1.1717  &  0.0     &  0.0     & 1.7620  \\
SkT6 & -1794.2   & 294.0   & -294.0   & 12817.0  &
          0.392   & -0.5     & -0.5     &  0.5    \\
SLy4 & -2488.913 & 486.818 & -546.395 & 13777.0  &
          0.8340  & -0.3438  & -1.0     & 1.3540  \\
SkI1 & -1913.619& 439.809& 2697.594 & 10592.267&
        -0.954536& -5.782388& -1.287379& -1.561421\\
SkI3 & -1762.88 & 561.608& -227.090 &  8106.2  &
         0.3083  & -1.1722  & -1.0907  &  1.2926  \\
SkI4 & -1855.827& 473.829& 1006.855 &  9703.607&
         0.405082& -2.889148& -1.325150&  1.145203\\
SkP & -2931.70    & 320.618   & -337.409& 18708.96&
            0.29215 &  0.65318  & -0.53732& 0.18103 \\
SkO &  -2103.653&   303.352&   791.674& 13553.252&
         -0.210701& -2.810752& -1.461595& -0.429881\\
SkO' & -2099.419&   301.531&   154.781& 13526.464&
       -0.029503& -1.325732& -2.323439& -0.147404
\end{tabular}

\begin{tabular}{|l|dddd|dddd|c|}
  Force & $b_4$ & $b'_4$ & $\alpha$& $\hbar^2/2m$ & $\theta_{\text{ls}} $& $\theta_{\text{CM}}$ &
     $V_p$  &  $V_n$ & source\\
\hline
SkM* &     65.0  &   65.0  &1/6 & 20.7525 &       0&      0&
              279.082&  258.962 &\cite{[Bar82]} \\
Z$_\sigma$ &     61.845&   61.845& 1/4  & 20.7525 &       1&      1&
              231.823    &  222.369 &\cite{[Fri86]}    \\
SkT6 &     53.5  &   53.5  &1/3 & 20.750 &       1&      0&
                202.526    &  204.977  &\cite{[Ton84]}   \\
SLy4 &     61.5  &   61.5  &1/6 & 20.73553&       0&      0&
              295.369&  286.669 &\cite{[Cha95a]}\\
SkI1 &   62.130&   62.130& 1/4  & 20.7525 &       0&      1&
              285.209&  291.384 &\cite{[Rei95]}\\
SkI3 &    94.254&    0.0  & 1/4  & 20.7525 &       0&      1&
         335.432 &     331.600 &\cite{[Rei95]}\\
SkI4 &   183.097& -180.351& 1/4 & 20.7525 &       0&      1&
              286.029&  310.832 &\cite{[Rei95]} \\
SkP &       50.0  &   50.0    &1/6 & 20.73&       1&      0&
              252.619&  236.237 &\cite{[Dob84]}\\
SkO &    176.578&-198.7490 & 1/4 & 20.73553&       0&      1&
                 253.771  &  269.942  &\cite{[Ben98a]}  \\
SkO' &  143.895&-82.8888 & 1/4 &  20.73553&       1&      1&
       256.095 & 258.122 &\cite{[Ben98a]} 
\end{tabular}
\end{table}

\narrowtext
\begin{table}
\caption{SMMC values of
$\langle Q^2\rangle$,
$\langle Q^2_n\rangle$,
$\langle Q^2_p\rangle$,
and $\langle Q_pQ_n\rangle$ (in $b^4$)  for $^{28,30,32,34}$Mg.
Typical error bar is $\pm$2$b^4$.
}
\begin{tabular}{cdddd}
\multicolumn{5}{c}{$sdpf$}\\
 Nucleus & $\langle Q^2\rangle$ & $\langle Q^2_p\rangle$ &
$\langle Q^2_n\rangle$ &  $\langle Q_pQ_n\rangle$  \\
\hline
$^{28}$Mg & 29.6 & 12.7 & 5.0 & 5.9 \\
$^{30}$Mg & 36.7 & 14.5 & 9.6 & 6.3 \\
$^{32}$Mg & 42.6 & 15.0 & 14.5 & 6.6 \\
$^{34}$Mg & 38.1 & 12.2 & 12.8 & 6.5 \\[2mm]
\multicolumn{5}{c}{$sdpf'$}\\
\hline
$^{28}$Mg & 37.2 & 13.8 & 11.7 & 5.9 \\
$^{30}$Mg & 30.0 & 13.3 & 8.7 & 4.0 \\
$^{32}$Mg & 42.8 & 13.5 & 18.2 & 5.5 \\
$^{34}$Mg & 68.1 & 13.4 & 35.7 & 9.5 \\[2mm]
\end{tabular}
\label{tabQ}
\end{table}

\clearpage
\widetext

\clearpage

\clearpage

\clearpage

\clearpage

\clearpage

\clearpage

\clearpage

\clearpage

\clearpage

\clearpage

\clearpage

\begin{figure}
\begin{center}
\leavevmode
\epsfxsize=16cm
\epsfbox{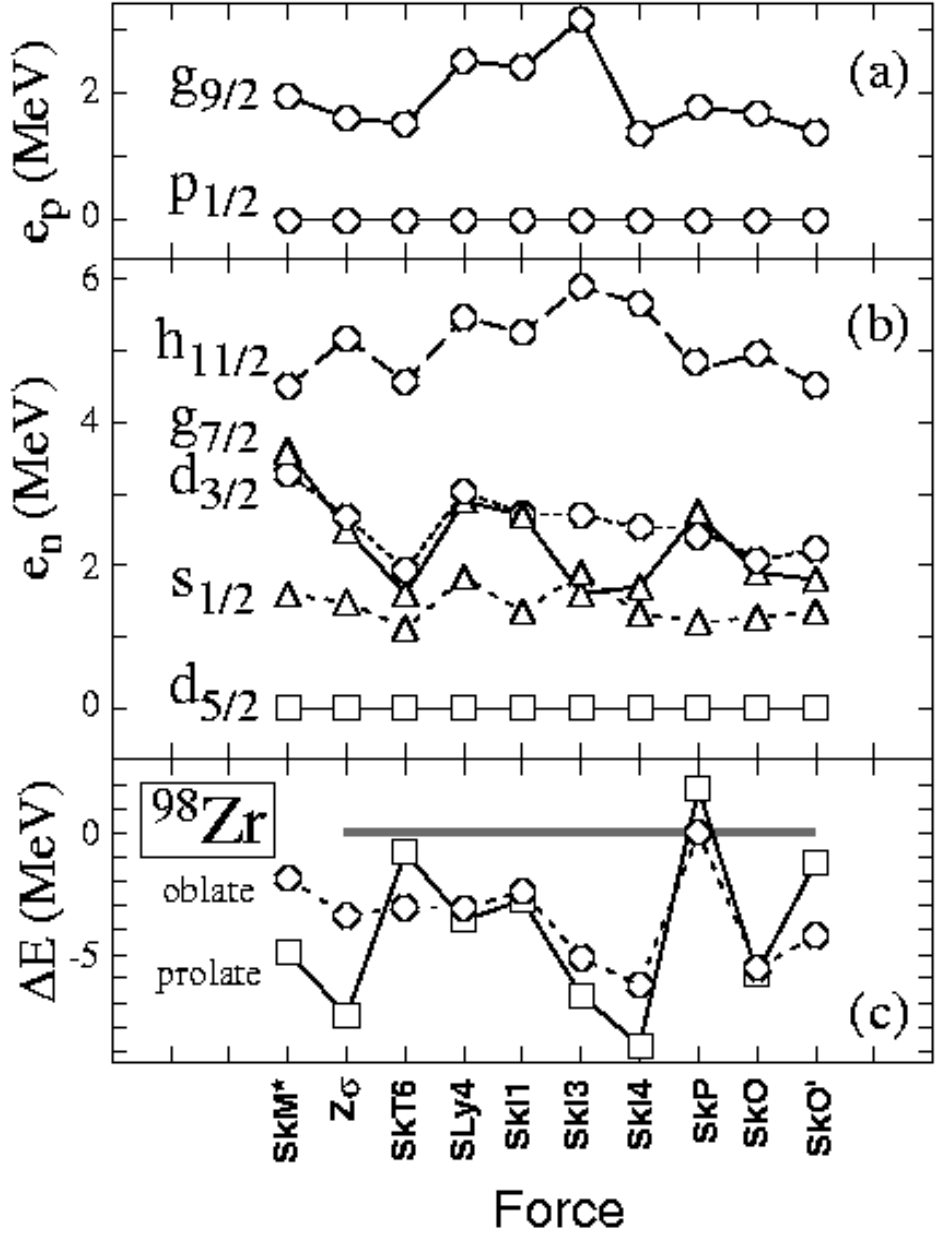}
\end{center}
\caption{Spherical neutron shell structure in $^{98}$Zr calculated
in several Skyrme-HF models.
 The single-particle levels are normalized
to the energy of the $p_{1/2}$ shell (a);
 spherical proton shell structure,
 the single-particle levels are normalized
to the energy of the $d_{5/2}$ shell (b);
the position of the deformed prolate and oblate minima
 with respect to the spherical HF state (no pairing)  (c)
}
\label{shell-Zr100}
\end{figure}
\clearpage

\clearpage

\clearpage

\clearpage

\clearpage

\clearpage

\clearpage

\clearpage

\clearpage

\clearpage

\clearpage

\clearpage

\end{document}